\documentclass[universe,article,accept,pdftex,moreauthors]{Definitions/mdpi} 

\firstpage{1} 
\pubvolume{1}
\issuenum{1}
\articlenumber{0}
\externaleditor{Academic Editor: Mark Henriksen}
\pubyear{2024}
\copyrightyear{2024}
\datereceived{14 March 2024} 
\daterevised{1 August 2024} %
\dateaccepted{3 August 2024} 
\datepublished{6 August 2024} 
\hreflink{https://doi.org/} %

\graphicspath{{./HiResFig/}{./Fig/}{./LowResFig/}{./}}

\Title{UHECR Clustering: Lightest Nuclei from Local Sheet Galaxies}

\TitleCitation{UHECR Clustering: Lightest Nuclei from Local Sheet Galaxies}

\Author{Daniele 
Fargion $^{1,2,}$*\orcidA{}, {Pier~Giorgio} {De~Sanctis~Lucentini} $^{3}$\orcidB{} and  {Maxim Yu.} Khlopov $^{4,5,6}$\orcidC{}}

\AuthorNames{Daniele Fargion, {Pier~Giorgio} {De~Sanctis~Lucentini} and {Maxim Yu.} Khlopov}

\AuthorCitation{Fargion, D.; {De Sanctis Lucentini}, P.G.;  Khlopov, M.Y.}

\address{%
$^{1}$ \quad Physics Department, Rome University 1, P.le A. Moro 5, 00185 Rome, Italy 
\\ %
$^{2}$ \quad  Mediterranian Institute of Fundamental Physics, 00047 Marino, Rome
\\
$^{3}$ \quad Physics Department, National University of Oil and Gas
Gubkin University, 
  65 Leninsky Prospekt, \mbox{Moscow 119991, Russia}; desanctislucentini.pg@gmail.com 
\\
$^{4}$ \quad Centre for Cosmoparticle Physics ``Cosmion”, National Research Nuclear University ”MEPHI”, \mbox{Kashirskoe Sh. 31}, Moscow 115409, Russia; khlopov@sfedu.ru\\
$^{5}$ \quad Virtual Institute of Astroparticle Physics, 75018 Paris, France\\
$^{6}$ \quad Institute of Physics, Southern Federal University, Rostov on Don 344090, Russia 
}
\corres{Correspondence: daniele.fargion@uniroma1.it}

\abstract{The ultra-high-energy cosmic ray (UHECR) puzzle is reviewed under the hints of a few basic results: clustering, anisotropy, asymmetry, bending, and composition changes with energies. We show how the lightest UHECR nuclei from the nearest AGN or Star-Burst sources, located  inside a few Mpc  Local Sheets, may explain, at best, the observed clustering of Hot Spots at tens EeV energy. 
Among the possible local extragalactic candidate sources,  we derived the main contribution of very few galactic sources. These are located in the Local Sheet plane within a distance of a few Mpc, ejecting UHECR at a few tens of EeV energy. UHECR also shine at lower energies of several EeV, partially feeding the Auger dipole  by  LMC and possibly a few nearer galactic sources. For the very recent highest energy UHECR event, if  a nucleon, it may be explained by a model based on the scattering of UHE ZeV neutrinos on low-mass relic neutrinos. Such scatterings are capable of correlating, via Z boson resonance, the most distant cosmic sources above the GZK bound with such an enigmatic UHECR event. Otherwise, these extreme events, if made by  the heaviest  composition, could originate from the largest bending trajectory of heaviest nuclei or from nearby sources, even galactic ones. In summary, the present lightest to heavy nuclei model UHECR from the Local Sheet could successfully correlate UHECR clustering with  the nearest   galaxies and AGN. Heavy UHECR may shine by being widely deflected from the Local Sheet or from past galactic, GRB, or SGR explosive ejection.} 


\begin{document}

\section{Cosmic Rays: A Brief History of a Wide Puzzle, in Search of a Model}



Cosmic rays (CR) are both neutral or charged particles.
Massless photons and gravitons are not usually considered as CR that are made by particles with a mass. Therefore, in this introduction, 
(1) we  first recall  the  relativistic  path of neutral particles and  their nearly straight trajectories in our sky or near gravitational objects;
(2) we also recall the main relativistic hadron interaction with cosmic photons leading to the nucleons scattering, to their loss of energy, and  to their bounded distances;
(3) we describe the  expression for the charged particles bending by cosmic magnetic fields, applying  them first to the  observed Moon and Sun shadows;
(4) we study the observed nuclear composition changes with the increase in their energies and their consequent different bending in their propagation within our Galaxy and nearby Universe;  
(5) we remind a first  probability estimate to validate our lightest nuclei model  correlating events with sources in the Local Sheet, since the last decade. We also recall the foreseen signals,  before their observations, of rare secondary traces and their associated multiplets along the few  UHECR spot regions along Cen~A.

\subsection{The Neutral Particle Deflections}%
Photons are massless and neutral particles. We can see them flying straight from their sources. Their bending in empty space is ruled only by gravity. Therefore, they offered, since Galileo in the last four centuries and since Maxwell in the last century,  a very sharp, fruitful multicolor astronomy. Their tiny gravity bending along masses was observed following the Einstein   $1915$ general relativity formula: 

\begin{equation}
   \Delta \varphi = \frac {4m}{b} 
\label{eq:DeltaPhoton}
\end{equation}
 In 
 the formula, the mass $m$ is $G\cdot M/ c^2$ size, while $b$ is the photon impact distance. This ratio is  twice the Schwarzschild radius of the body over the minimal impact photon skimming distance, usually the star radius. This tiny deflection along our Sun's edge was historically observed  only in  $1919$.
 
The neutral particle  with mass trajectory follows different geodesics. Therefore, their bending is more complex. At nonrelativistic regimes, they follow a path like the Newtonian one, but  at relativistic regimes, they are almost identical to massless particles in general relativity theory.

\begin{equation}
   \Delta \varphi = \frac {2m}{b \cdot \beta_{\infty}^{2}}{(1+ \beta_{\infty}^{2})}  =  \frac {4m}{b}  \frac {(1-1/2\gamma_{\infty}^{2})}{(1-1/\gamma_{\infty}^{2})} \simeq \frac {4m}{b}  \left(1+ \frac {1}{2\gamma_{\infty}^{2}} \right) 
\label{eq:DeltaParticle}
\end{equation}

Therefore, relativistic neutral particles might offer a new sharp astronomy~\cite{Fargion:1981ge}. Their mass will introduce a small or large time delay with respect to the massless ones, allowing them to disentangle their tiny or large mass~\cite{Fargion:1981gg}.
Neutral particles with mass,  known in nature, are either the lightest stable neutrinos or the unstable neutrons. Many other unstable mesons, such as the pions or the unstable lepton, are too unstable to play any key  astrophysical role. 
Neutrinos are weakly interacting particles that are very difficult to observe; their signals are  polluted by cosmic ray secondaries and by their decay, the muons, and the atmospheric neutrino noises. The neutrons are quite unstable and cannot fly from cosmic distances, excluding the eventual EeV or hundred EeV energetic ones in our galactic~halo.
    
    Therefore, excluding the few highest energy neutrinos  possibly observable in the largest underground detectors~\cite{IceCube:2022der},  SUSY neutral astronomy~\cite{Datta:2004sr} should wait for better detectors to make a guaranteed  astronomy~\cite{fargion2004tau}. Also, the solar  neutrinos and  nearest  historic supernovae, SN 1987A, occurred nearly half a century ago, they offered the first fundamental neutrino astronomy signatures and their flavour-mixing nature. 
    
    Charged particle  air showers made by ultra-high-energy cosmic rays, UHECR, are a much more  abundant  signal in the sky.  

Indeed, since  the first decades of the last century, such charged CR have puzzled both physics and astrophysics. Both for their acceleration and composition as well as for their~sources.

The CR charges and  their effective random cosmic magnetic bending make it difficult to recognize  the original CR  source in the~sky. 

It is like having a smeared glass in front of our eyes.  The source distance, their CR energy, and the nuclei charge  make most of the distortion on the source image,  its profile, and sharpness.

CR have a long history linked to nuclear physics discoveries.  
In the 1920s, it was noted that Earth emits very little radioactivity at ground level but, while at high altitudes, thanks to balloon detectors flying kilometers above sea level, it was found that the radioactivity  increased significantly. A huge factor, up to 30 times for leptons (mainly electron pairs) and about a thousand times for protons,  is He or nuclei at the edge of Earth's atmosphere. 

The atmosphere and the bending made by geomagnetic fields shield us from this dangerous radioactivity.

Our Sun, as well as other galactic and cosmic sources, could send us such charges, proton or nuclei, expelled, for instance, by magnetic solar flares at GeV energies. At higher energy, CR might be ejected by supernovae, SN explosions, or blazing jets by  Gamma Ray Bursts, GRB, or longer life Soft Gamma Repeaters, SGR; on a larger scale, these jets arrive from  Active Galactic Nuclei, AGN, around the largest black hole, BH. All or most of such sources often have external accretion disks feeding the jets. The UHECR' most energetic events are very probably  connected with the tidal disruption of a binary Neutron Star, NS, system,  or the binary black hole, BH, and NS, explosive jet emission. These beamed ejections, once spinning and precessing in collimated axis to us, appear as sudden gamma, X  brightening  called GRB or SGR, often inside SN Remnants, SNR~\cite{fargion1999nature}. A rich galactic population cluster might contain many  of such  explosive events; therefore, they are often  called star burst galaxies, SBG, a kind of AGN active region.

Therefore, we have used, in the last century, a simple and generic \textit{cosmic ray 
} term, following pioneers Robert Millikan and Bruno Rossi's visions. However, their origin might be solar, galactic, or extragalactic and  therefore also cosmic, as their name suggests; see~\cite{Grieder:2001ct, SOLDIN2024102992}.

\subsection{UHECR Cut-Off and Local Sheet Galactic Sources}%
The most energetic UHECR particles, above  $10^{18}$ eV, or Exa-electron-Volt (EeV) energy, were expected to mainly originate from 
truly far cosmic sources. 
Actually, the UHECR cannot fly more for two reasons, the
magnetic field bending and their spiral trajectory,  as well as the additional photopion 
opacity, named after its authors Greisen, Zatsepin, and Kuzmin \cite{greisen1966end,1966JETPL...4...78Z}, 
the (GZK) cut-off. 
 This GZK limit constrains the UHECR arrival within a small, nearly $1\%$ of the cosmic radius
size, about 40 Mpc. The UHECRs are so energetic so as to be (for the proton nucleon or for the
lightest nuclei) almost undeflected, leading in principle to a fruitful source identification.  
They are so rare but so energetic so as to be observable on the ground by their wide size, 
tree-like air shower. Their secondary tails are recorded in huge arrays spread over thousand-kilometer areas.  If UHECR are protons, they should reach us quite directly from Virgo, which is 20 Mpc away. Virgo is the most luminous Infrared cluster containing a few thousand galaxies, the brightest source shining in infrared within our narrow GZK Universe. This UHECR Virgo source was not found in the last two decades. 
Its absence (or negligible signal) early in the year $2007$ and its missing in late Auger and TA records had been the main argument forcing  the need for a lightest nuclei model
 \cite{Fargion:2008sp}; a model able to screen UHECRs and stop the  expected Virgo signal.
 Indeed, luckily, there was an additional constraint or cut-off, mostly for the lightest UHECR  nuclei but not for the protons: it is the photo-nuclear disruption or disintegration due to  the GDR (Giant Dipole Resonance). Indeed, infrared or cosmic photons might produce  effective scattering on the UHECR lightest nuclei propagation, splitting their masses.  This cut-off is made by the photo-nuclear disintegration that reduces and breaks  these nuclei severely (such as D, He, Li, Be) in their path. Their reduced propagation path allows  them only a smaller size Universe, as small as $0.1\%$ of the cosmic radius. 
A very local 3--4 Mpc cut-off occurs above a few tens EeV UHECR energy. 
Heavier nuclei may be  less fragmented and less constrained; they are nevertheless more charged and more  deflected, leading to a less effective track and a less meaningful directionality.  The lightest UHECR nuclei's main role nowadays has been, step by step, widely  accepted. They are abundant in the Auger and TA UHECR sky at few tens EeV energy, where  the main UHECR event clusters, the hot spot, occurred. They must be formed within the so-called Local Sheet, which is a few Mpc wide, ruled by the so-called Council of Giants~\cite{McCall:2014eha},
a place populated by a few large galactic sources; see below in Section \ref{7}.
Also, our nearest galaxy, Andromeda, or even our Local Magellanic Group, or a few rare  brightest (galactic) gamma sources, could be the sudden UHECR source, ejecting in the near  past, at a few EeVs energy. Indeed, the UHECR ejection may be very rare, once every thousands of years, and very beamed, making their blazing in our galaxy quite inhomogeneous and,  as we shall see, possibly feeding the observed wide Auger dipole anisotropy  at several EeV.



\subsection{The Charged Cosmic Ray Bending and the Solar and Lunar Shadow Shifts}
The PeV (Peva--electron--Volt energy range, about $10^{15}$ eV) CR, or those at energy just below, should be bounded within our galaxy because their spiral gyro-magnetic radius is much below the galactic sizes.  UHECR at few or higher EeV energy are possibly both galactic as well as partially extragalactic.

Usually, we cannot see them by eye. But the abundant MeV-GeV energy solar CR flares are clearly visible: they shine on Earth's atmosphere like auroras. The largest, low-energy ones could be very dangerous for orbital satellites, perhaps even lethal for astronauts leaving from the Moon or  making long interplanetary flights.

Nevertheless, we may test  this CR bending at the nearest distances for the solar and the lunar shadows: The theoretical formula for Lorentz force and its consequent  Larmor bending radius is well known.
  
  The ratio between the characteristic distance in flight (under the main bending force) and the gyro-radius offers the approximated deflection angle. For a proton, the relativistic Larmor radius $R_L$ is 

    \begin{equation*} R_L = 33360~\text{km} \cdot (E/TeV) \cdot (B/Gauss)^{-1}\end{equation*}
  
Let us  remind readers here  of the  ideal case for a coherent bending of the geomagnetic field for  the Moon's  shadow.
  
  Earth's magnetic  dipole field decays as a cubic law with distance. 
  Therefore, the deflection is being mostly ruled (while being  displaced to the west side, because nucleons are positives) by the nearest geomagnetic dipole in a very narrow Earth size  by an approximated formula as

  \begin{equation}\label{angleMoon}
\delta^{
}_{\mathrm{G_{Moon}}}=1.27^\circ\left(\frac{Z}{Z_{p}}\right)\left(\frac{E}{T\,\mathrm{eV}}\right)^{-1}\left(\frac{D}{4000~\text{km}}\right)\left(\frac{B}{0.2 G}\right)
\end{equation}
  
  These bending are used to calibrate  TeV-PeV cosmic ray resolution in the CR detector array. The solar shadows suffer a comparable bending, even if one includes an  additional deflection  that is related to the solar interplanetary influence in the far CR flight: 

    \begin{equation}\label{angleSun}
\delta^{
}_{\mathrm{G_{Sun}}}=1.15^\circ\left(\frac{Z}{Z_{p}}\right)\left(\frac{E}{T\,\mathrm{eV}}\right)^{-1}\left(\frac{D}{1.5 \times 10^8~\text{km}}\right)\left(\frac{B}{5 \mu G}\right)
\end{equation}
  
 Different incoherent bending  expressions, the random ones,  are useful and are considered later in the article.

  The abundant nearly homogeneous rain of such TeV (Tera-electron-Volt) CR, up to tens or hundreds TeV ones, are marked by their  Sun or Moon disk dark opacity. These displaced shadows nearly overlap with the optical ones at tens of TeVs. These tests are  fundamental tools for our understanding of CR bending and for the CR array \mbox{angular~precision.}
  
  At the highest UHECR energy edges,  these deflections are even more negligible. However, their UHECR flux is too diluted and rare at a hundred PeVs to allow their statistical imprint or shadows,  even by the present, largest, ground array. 
  Nevertheless, UHECR in the last decade showed, thanks to their rigidity,  some new meaningful anisotropies and clustering by the largest  array detectors (such as Auger and TA, Telescope Array), signals possibly of  an astrophysical nature discussed in the present article.

Let us remark that a full understanding of CR is a multi-parameter puzzle that is quite complex to solve. The collective or random bending of the planetary, galactic, or intergalactic  magnetic fields leads to such a mix of data.
Also, the delayed timing of their arrival, with respect to faster photons, plays an additional mixing and confusing role. 
Finally, the nuclear photo-disruption or the photopion opacity could bound the UHECR horizons in a quite narrow cosmic Universe. The nearly monochromatic composition of each of the different UHECR at different energies allows a better understanding of their puzzle behavior, as described below.

Historically, the CR signals have extended from  Mega electron-Volt,  MeV, range  for hadrons up to hundreds EeV for the UHECR nuclei~\cite{PierreAuger:2007pcg,COLEMAN2023102819, aloisio2023ultra, caccianiga2023update, AbdulHalim:20232A, ANCHORDOQUI20191}.
All the observed CR energy  ranges span  nearly $14$ orders of magnitude, with a flux (the rate number) spanning nearly $19$  orders of magnitude; see Figure~1. Their flux decreases very rapidly with their energy growth. The UHECR composition evolves much more slowly with their energy increase.


\subsection{The UHECR Composition Changes with the Energy in the Nearby Universe}%

The UHECR understanding is still contradictory among many  models that have evolved in the last years~\cite{COLEMAN2023102819, caccianiga2023update, ANCHORDOQUI20191}. 
Two decades ago, the first discovery  by Hires~\cite{BERGMAN200719} of an apparent UHECR cut-off at the energy of $6 \times 10^{19}$ eV was generally understood as being related to the  proton--photopion, GZK cut-off. 
This evidence led most authors to assume that the proton is the main UHECR courier. As we mentioned, in the proton case, the GZK observable volume extends nearly up to 40--100 Mpc. 

The early negligible galactic signature led most authors to claim a definitive extragalactic UHECR nature. Also apparent is the correlation of a few tens of UHECR events with the Super-Galactic  plane~\cite{PierreAuger:2007pcg} that made most convinced of the proton courier role bounded by the GZK cut-off. Today, because of the precise air shower slant depth signature since 2017, the signature favoring the lightest or light nuclei, very few authors are still considering  proton contribution as the main one in UHECR.  

Since the first Auger results in 2007~\cite{PierreAuger:2007pcg}, some authors  had quite different concerns~\cite{Fargion:2008sp}. 
They mainly wondered about the absence of the Virgo cluster within the proton GZK volume in Auger data. This dominant presence had to be expected if  UHECR were indeed protons.  
The first Auger  statements on the first  anisotropy results~\cite{PierreAuger:2007pcg}, showed only a single UHECR cluster along our nearest  AGN, Cen~A, as a   hot spot.
A new interpretation has therefore been imagined and suggested, favoring not protons but the lightest nuclei (well before their composition had clearly been revealed)  for these UHECR~\cite{Fargion:2008sp}. Because of the lightest nuclei fragility, they had to be more bound  by the earlier photonuclear disruption in a very local environment of a few Mpc. A distance  nearly ten to twenty times smaller than the same small GZK cut-off due to photopion reactions. This nearer distance includes our nearby AGN Cen~A, as observed in Auger. But it also excludes the more loud  rich and far Virgo cluster inside the GZK volume.

As mentioned, this  interpretation was needed to solve the unexplained  Virgo cluster  missing, located at 20 Mpc, in Auger  UHECR data.
 Several years later, a few additional nearby AGN or star burst source candidatures  appeared in Telescope Array, TA, and Auger maps. All of them, M82, NGC 253, and Cen~A, are within 3--4 Mpc volume.   These three  candidates are somehow ruling the present UHECR anisotropies and are all located within the Local Sheet galaxy structure.
 
Recently~\cite{aab2017combined}, Auger measurements of UHECR air shower profile indicated a nuclei mass composition  that becomes heavier (than a proton) with their energy increase, giving more support to the lightest nuclei model \cite{Fargion:2008sp}.

In present  Auger composition data, the protons rule mainly below ten EeV, while the lightest (He-like ones) nuclei rule at   1--4 $\times 10^{19} $ eV range; other light nuclei are present at  4--8 $\times 10^{19} $ eV. Finally, UHECR spectra showed the  presence of heavy nuclei with increasing energy, between 8--20 $\times 10^{19} $ eV, just near the GZK limit.   
 These results  make the UHECR  proton models no longer acceptable. Indeed, these UHECR  proton models would also lead to GZK secondary tau neutrino at EeV energy, which had to shine in Auger and TA, as horizontal--upward air showers. These expected signals are absent in the last decade's data.
 
 The expectations for the lightest nuclei model are consistent with the data.
  
At present, the question of the  GZK cut-off itself in UHECR spectra as a real cosmic GZK opacity on the nuclei or an intrinsic bound linked to the composition limit and the UHECR acceleration process is still debated.

The  updated model, based on the lightest nuclei,  has been extended  in the last decade and is reviewed nowadays~\cite{FargionICRC2023PoS395} .
 As mentioned, after the Cen~A candidate, M82 formed the second hot spot map.
  A decade ago, in 2015, the third NGC 253  UHECR source was recognized within Local Sheet volume~\cite{FargionICRC2023PoS395}.
The lightest and  also heavier nuclei can mainly originate in this very  narrow    Local Sheet~\cite{McCall:2014eha} volume. 

As we discuss later,  an additional few galactic UHECR sources, such as LMC,  could  contribute to the ten EeV dipole anisotropy observed in the Auger. 
The idea of a very local UHECR, in the last few years, is no longer unique. Indeed, a few authors reached very recently the same idea of local (Mpcs) UHECR models, giving key relevance to the star burst as the ideal track for UHECR sources~\cite{Taylor:2023qdy,Biteau:2021pru}. Their local models are partially coincident with the earlier ones~\cite{FargionICRC2023PoS395}. 
Therefore, the models   are complementary.

\subsection{UHECR: Probability to  Correlate Hot  Spots to Their Local Sources }%
The statistical probability in the lightest nuclei model to fit the present observed  hot spot might be estimated assuming  the key note on the absence of the  (expected) Virgo source.
Therefore, the UHECR originated from the ideal Virgo cluster, located within the photo-pion GZK volume, could not be a nucleon.  
Indeed, the composition  favoring the lightest nuclei had been accepted by the UHECR air shower's better profile study in Auger, but only since a more recent discovery~\cite{aab2017combined}. 

In the ideal case of the lightest nuclei model,  one had to foresee and recognize the few largest nearby galaxies as the ideal hot spots. Indeed, as shown  in nearby Local Sheet giant sources (see Figure~8) these  are the best candidates.
  Therefore,  by correlating these five main galaxies, as labeled in Figures~8 and 10  by red rings, with future UHECR clustering, one may estimate the ideal correlation. Assuming a solid angle for each of the five sources (Cen~A, NGC 253, M 82, M31, Maffei)  as large as a beam size $\pm 18 ^\circ$, or $2.5\%$ of the whole sky view, then the probability $P_{5}$ to find their five directions as (which have partially been observed) five~hot spots in the Auger-TA clustering is simply

\begin{equation}\label{eq:Prob5}
   P_{5} = (1/40)^{5} =9.8 \times 10^{-9}
\end{equation}

      This was not the real case because, in $2008$, the Cen~A presence as a first hot spot was already known. The second nearby source in the North sky was later discovered by TA. 
      
         Therefore, a possible prediction that was allowed by the lightest model in $2013$ was to foresee the presence of the next three candidates: the Andromeda, NGC 253, and Maffei Galaxies.  In those years, the Local Sheet role was not yet widely known \cite{McCall:2014eha}.  The Local Group is  small and contains a couple of galaxies. The Local Sheet is wider,  including the Local Group and a dozen sources. The Super-Galactic plane does contain the Local Sheet but extends to GZK volumes that are not allowed for the lightest nuclei. 
          Therefore, a decade ago (see review in~\cite{FargionICRC2023PoS395}), only the possibility of a new  (still unnoticed)  Star Burst, NGC $253$, was foreseen, as well as the eventual nearest Andromeda.
          
          The Bernoulli probability $P_{2}$ that was foreseen, {\it a priori} by chance,  within the last three main Local Sheet sources, at least two, is
          
\begin{equation}\label{eq:Prob2}
   P_{2} = 1.828 \times 10^{-3}
\end{equation}

 If the authors had been aware of the Maffei role in the Local Sheet, whose clustering is now partially present in  Auger data, then its early suggestion would imply a much better Bernoulli probability $P_{3}$ to find by chance three out four main candidates: 

\begin{equation}\label{eq:Prob3}
   P_{3} = 6.09 \times 10^{-5}
\end{equation}
          
 All these four sources are correlated with the five main galaxies in the Local Sheet, as shown in  Figure~10.
 
   The ability of the lightest nuclei~\cite{Fargion:2008sp}, based only on the Virgo absence,  to foresee~\cite{aab2017combined} the UHECR composition was itself a remarkable success.
   
   The model also suggested the presence of twin multiplets associated with the Cen~A~\cite{FARGION2011111} due to the UHECR fragmentation. These multiplets were soon later observed~\cite{abreu2012search}; also more recently, a third multiplet was found pointing to NGC~253~\cite{Fargion:2023yiy}, an extremely remarkable correlation.
  Moreover, the Local Sheet galactic masses distributions in our celestial coordinate are not symmetric. 
   The detailed study of the Local Sheet mass distribution in the Northern and Southern  hemispheres shows some overabundance: North hemisphere mass densities, at 2--4 Mpc distance \cite{Biteau:2021pru}, are a few times larger than the South ones.  Comparable to the UHECR  rate asymmetry observed  in North and South UHECR spectra, there is a more abundant flux of UHECR at TA array, North sky, with respect to the less flux in Auger, South sky.  
The probability that the two asymmetries occurred by chance at similar ratios and in the same direction is negligible. UHECRs from Local Sheet sources explain the Auger-TA spectra puzzle.
   
 In addition, we consider the probable role of a few UHECR galactic source candidates that might be able to partially fit and complete the surprising Auger  dipole anisotropy.

\section{ UHECR: Lightest to Heavy Nuclei Inside a Bounded, Smeared, Astronomy}%

Regarding the UHECR mass composition, the average shower depth distribution $X_{max}$, (an air shower shape growth), measured by Auger and TA \cite{aab2017combined},
shows that the UHECR nuclei mass  becomes heavier with its increasing energy.

This occurs clearly above $2 \times 10^{18}$ eV energy. The measured fluctuations of $X_{max}$ point out a small mass dispersion at energies above the ankle. It indicates one main nuclear carrier species at each energy range, as shown clearly in Figure~7.

For UHECR composed only of protons, the random magnetic field bending at GZK edges would be a marginal one, just a few degrees: their consequent  signals should point almost sharply toward  their defined source directions.

In the case of the lightest nuclei (D, He, Li, Be), the trajectory bending should be a little wider,  by  an angle of about ten degrees~\cite{FARGION2011111}, as it has been observed in hot spots. 
The lightest nuclei are located in the nearest Universe with a few Mpc radius due to their nuclear fragility.

Heavier nuclei (C, N, O) or the heaviest ones (Ni, Co, Fe) do not suffer much of the photo-nuclear disruption.
However, these heavy nuclei are more and more deflected and spread, arriving from  a volume nearly comparable to or smaller than the GZK one. Their smeared contribution may induce only isotropic noises. 

All proton to light and heavy nuclei compositions are  present in different UHECR energy windows of the last decade's data. 

The composition changes smoothly with energy, from the lightest to heavy: they could also play different roles in the spectra shape  changes; see Figures~2 and~7.

The Virgo cluster is almost absent in the UHECR map; see Figure~3.

Instead, it should be prominent (if UHECR were protons), because it is the dominant cluster in the GZK volume.
It should be quite collimated for such a proton courier. 

 Therefore, UHECR protons cannot play any relevant cosmic role at several tens EeV at energies where anisotropy and the hot spot emerged.
 
 This was and is the first, main  base~\cite{Fargion:2008sp} of the lightest nuclei model.

We underline how, at tens of EeV, the main  clustered  UHECR signals are probably ejected by a very few  nearby  AGN, or Star-Burst galaxies, such as those few in the Local Sheet \mbox{nearby Universe.} 

At several EeV, UHECRs could also  be polluted  by a few unexpected galactic microquasars or SNR  sources, partially feeding  the established wide dipole clustering discovered in the Auger sky; see Figure~5. 

For very few  remaining and unexplained UHECR events,  we suggest the largest coherent bending of heaviest nuclei eventually located in the Local Sheet and  also in nearby galactic sources; see Figure~6. 

For the most puzzling UHECR events, we recall  a different model based on cosmic relic neutrino halos, dark matter particles with small mass. 
The relic neutrinos, while being hit by ZeV (Zeta, $10^{21}$ eV)  UHE neutrinos in Z boson resonance energy, can  produce, in decay,  UHECR nucleons as secondaries. These nucleons are ejected within our surrounding Mpc in a hot dark halo.  This process is able to overcome any GZK  cut-off, even from the most distant cosmic  edges~\cite{fargion1999ultra,Weiler:1997sh}. 
 
However, the \textit{lightest UHECR nuclei and rare heavier nuclei}  model~\cite{Fargion:2008sp} from the nearby Universe with few defined candidates along the Local Sheet remains, as discussed in this article, the simplest and most satisfactory updated option~\cite{FargionICRC2023PoS395,Fargion:2023yiy}. 
 
As we mentioned, CR are mainly charged particles (excluding, for the moment, photons, our basic astronomy signals).
 Massless gravitons and the nearly massless ultra-relativistic neutrinos could also lead to a new astronomy. 
So far, CR do not offer astronomical maps as defined and good as photon ones.  UHECR made by neutrons, being undeflected by magnetic fields, could be ideal but are difficult to produce. They are not yet observed. So, the UHECR, even charged, could represent a fruitful  smeared new astronomy, as most authors have been expecting for decades.

 CRs are mostly hadrons, because lepton companions, also present in a small amount as electron pairs, suffer more severe energy losses during their flight. Indeed, their primary electrons are more constrained to their sources, within a few parsecs, much nearer than the  protons or the UHECR nuclei. Leptons, which lose   their energy in magnetic fields or by photon scattering much faster, are  brighter (as by radio, optical, and X-rays) than hadrons. Lepton secondaries may better trace the UHECR origination and tails in the space. 
 
 These secondary CR, mainly electron pairs,  sometimes reveal their correlated UHECR primary hadron jets, creating observable giant  astrophysical tails and tracks: observable radio-X and gamma, SN shells, hour-glass jets from AGN, or microquasars beams. 
 
 These UHECR particles are ejected by shock waves, in SN, or in the tidal collapse, as well as in the accreting disks around the black hole, BH, or NS, feeding their jets.

\begin{figure}[H]
\includegraphics[width=10.5 cm,height=10.0 cm]{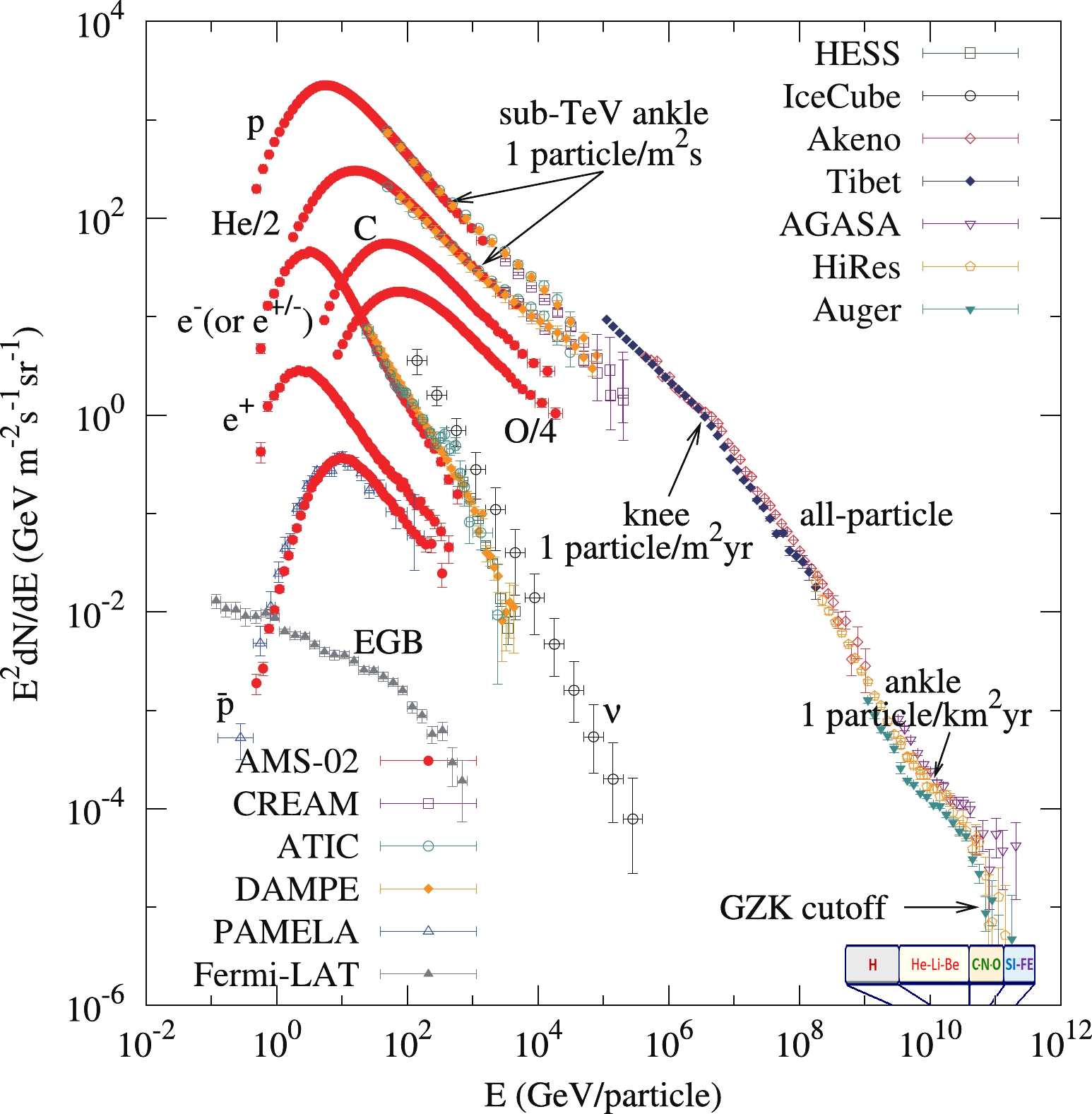}
\hfill
\caption[]{
The 
 updated Cosmic-ray spectra for most lepton and nuclei, based on the remarkable review in \cite{Piazzoli_2022}. At the bottom right, the most probable composition intervals, from~\cite{AbdulHalim:20232A}, are highlighted and expanded to include the lightest nuclei ($D, He, Li, Be$), as explained in the text.\\
\label{CR_SPECTRA_LIU}
}
\vspace{-10pt}
\end{figure}

So, electron pairs and even rare  antiprotons can reach us  as secondaries of CR and UHECR scattering in flight within dense or dilute space; see Figure~\ref{CR_SPECTRA_LIU}.

The highest-energy UHECRs could be less bent by magnetic fields due to their ultra-relativistic energy and their greater inertia, which causes them to fly along straighter trajectories.

However, their composition evolves from lighter to heavier nuclei as energy increases~\cite{aab2017combined}. 
These changes could lead, once again, to  a bent, fuzzy, and inconclusive high-energy UHECR astronomy. 

Apparently, the lightest nuclei are the only ones among the UHECR with some ability to keep enough directionality, leading in the low tens of EeV to a large dipole anisotropy and some north--south celestial asymmetry, clustering  in more narrow \textit{hot spots} at several tens of EeV, forming rare multiplet trains of correlated events. 

This possibly leads, for protons and the lightest nuclei at several EeVs, to a very smeared CR map, such as the Auger dipole one.
Then, at ten EeVs, the lightest nuclei may also feed  the anisotropic UHECR maps;  see Figure~5.
Later, at a few tens of EeV, a few hot spots arise from the lightest nuclei. Finally, around the highest energy edges, more heavy and charged nuclei lead, again, to a very smeared,  isotropic map.

  Supernovae  explosions could accelerate CRs up to PeV energies. Mainly, NS,  BH, star burst galaxies, or  AGN jets could be the ideal accelerator for higher UHECRs. 

 Such CRs, UHECRs, and their GeV electron pair tails are observable in radio waves. These UHE electrons in narrow beams may also blaze to us as bright X-ray or gamma radiation.
 
 These jets are orthogonal to the accretion disk, precess, and radiation. 
 While their cone beam is on axis with us, they can appear as an explosive, sharp variable burst of gamma rays, a GRB~\cite{fargion1999nature}.
 
 Their flashes are observed on axis as  slow flares  from AGN (largest BH)  or fast from GRB  (smallest BH) or NS. The variable and somehow sudden  appearance of Soft Gamma Repeaters (SGR), or Gamma Ray Bursts (GRB)  could be made by such a long life, narrow precessing  gamma jet~\cite{fargion1998inverse} by the Inverse Compton of electron pair at GeV-TeV energies.  These electron pairs may annihilate at a hundred-tens MeV energy. 
 
 The associated UHECR acceleration, ejection, their smearing at different energies, epochs, and distances, and the different nuclear compositions made the discovery of UHECR sources such a difficult puzzle to solve.

\section{UHECR: Persistent or Transient? Beamed or Spherical? The North--South Asymmetry}

Different astrophysical events have different lifetimes. Usually, the brightest ones have the shortest lifespans.
While photons, neutrinos, and gravitons fly straight and simultaneously at the speed of light, CR and UHECR bend into arcs or spirals following longer and more diluted trajectories. This occurs because they are charges that pass through random cosmic magnetic fields.
So, even if UHECR originated by a prompt  burst of a few seconds or hours, their arrival could still be delayed by thousands or millions of years if reaching us from cosmic distances, or in shorter times if within our galaxy.

Eventual exotic sources of UHECR, by their decay,  may also not be stable: any relic heavy particle from the primordial Universe could decay or disappear via pair annihilation, becoming a source of cosmic UHECR events: however, the recent discovery of anisotropy and clustering, possibly correlated to Local Sheet sources, does not promote such \mbox{exotic models}.   
 
Therefore, we should better consider  the timescale of the UHECR source activity associated with astrophysical known objects such as SN, microquasars, AGN, and star~bursts. 
  
A main question arises: How long could  the transient events that are ejecting UHECR survive? 
They may be the explosive, week-long events such as supernovae, or a more rare and sometimes explosive binary NS-NS or NS-BH tidal disruption, blazing in beamed jets. 

Are the UHECR emitted by their relic persistent, decaying, spinning, precessing jets~\cite{fargion1999nature}, which appear to us as GRB or SGR, once collinear with their axis?  Otherwise, are they instead the product of a single-emission fixed jet at the peak of the GRB or an AGN \mbox{activity flare?} 

The idea that GRB and AGN are sources of UHECR is not new and is quite widespread.%

Such an event could also be fed by a longer-life relic binary accreting disk system, like the celebrated microquasars SS433. These jets could be active longer than their starting, explosive, birth,  leading to SN remnant (SNR)  and a Soft Gamma Repeaters (SGR). 
Their core may contain  a few or several solar masses black holes . 

Other larger extragalactic jets such as quasars or AGN are fed by their million mass BHs. 
Their  jets are flaring, modulated by cannibal star collapse while precessing and blazing  in quasi-periodic modes. %
Rare GRB or NS-NS tidal collapse could explode at peak output with a beaming jet, pointing to us, even partially off-axis, while emitting gravitational waves. This very peculiar geometry and rare signal was already identified in~2017. 

The main difference to underline is that UHECR jets are beamed, while their gravitational wave, GW, emission is almost spherical. This implies that only very rarely both the near distance (essential for GW) and the beamed jet collimation to us (basic for  UHECR) are taking place. 
This is the same reason for the surprise of a lucky, rare, event, such as a short gamma burst (a conical beamed jet) and the associated GW (spherical one)  observed once in 2017.

Moreover, the time delay and the deflection of the UHECRs would make it more difficult for any future correlated UHECR-GW event to be discovered.
   
Any transient jets, while spinning and multiprecessing, could appear  trembling, as a variable X, by multipeak  gamma burst or SGR, being pointed on-axis and off-axis to us. %
Therefore, such  repetitive beaming might play a role in rarer, nearest UHECR multistart repetitive events, such as the multiplet ones.
The hard gammas are often provided by electron pair scattering by the electron--photon inverse Compton scattering, ICS,  onto  self-radio-synchrotron or thermal black body photons~\cite{fargion1998inverse}.  
However, the UHECR blaze, even though short in its starting time, will be bent and delayed. Its arrival by different curvature trajectories, due to the  galactic and cosmic  magnetic field effect, will spread its arrival signals. This occurs for different energies, magnetic fields, UHECR masses, and charges in consequent diluted  time flights along their different trajectories. %

  These AGN, or even their microquasar copies, might be able to accelerate particles to UHECR at their peak activity at birth. The NS-NS tidal disruption occurs  during the star-NS, NS-NS, or NS-BH nearest encounter, leading to GRB or late SGR. 
The star (or NS) is feeding a disk toward the BH companion, which is the source of an orthogonal jet  engine. 
  Such a star-BH jet is originally made by hadronic nuclei and their surrounding electron pairs. 
  Their secondary lepton fragments observable along the jet are scattered on thermal and synchrotron photons, feeding powerful gamma jets. 
  These NS-NS (or more rare NS-BH) collapse events occur on average only once every $10^3$--$10^5$ years in a galaxy like ours. The coherent (or random) spread of their UHECR flights in galactic fields may dilute their short flash into a few hundreds (or thousands) of years, as a train of events, observable in clustered or multiplet traces. Such multiplet signals have been rarely observed, as we discuss below. %
  Because of the common narrow beaming of jet angles and because of their poor rate, we expect only a few of the recent powerful local jets to be blazing to us now by their UHECR tail: around nearly $10^5$--$10^3$ years, the past galactic binary NS-NS encounters or NS-BH   narrow collapses, and only a few might be beamed to us at their birth. Only a few could show their present records by their  UHECR.  Possibly, the few brightest active jets  in the nearest star burst galaxy or  AGN and even the few galactic microquasar such as SS433, or brightest NS pulsar as Crab and Vela.
  
  Could UHECRs then also be galactic? 
  
  In this case, the question about the absence %
  of our Galactic Center in UHECR emission has to be answered.
  We remind readers of the orthogonal twin-hour-glass gamma fountain, first revealed by EGRET decades ago and then better by Fermi gamma satellites.  Our Galactic Center (GC) is not active at present, and anyway, it was  flashing and beaming as a fountain far from the Milky Way plane. 
  Moreover, GC  is  not pointing to us. Many other SNR-associated jets might be generally pointing elsewhere. 
But some, the brightest ones, may be old GRB remnants that had been the source of EeV's UHECR, reaching us as signals, diluted over time. A change in spectra in UHECR occurs at $5$~EeV. It could  be due to the injection of extragalactic UHECR (Local Sheet sources), or better,  to a change in UHECR composition from proton to lightest nuclei.
  
  Extragalactic sources ejected from the nearest AGN or star burst have a longer lifetime and wider beam spread but are  very rare. Moreover, we remind readers that the lightest nuclei UHECR composition  from  Auger and TA records   make the extragalactic sources (AGN, star burst, or nearest galaxy) located only in a few Mpc distances.
  
  The GZK cut-off for photo-pions for nucleons as protons reduces the  UHECRs survival distance to 40--100 Mpc or below~\cite{greisen1966end,1966JETPL...4...78Z}. 
  The more fragile and earlier (less energetic)  photo-nuclear disruption for the lightest nuclei,  due to the giant nuclear resonance,  will limit its flight to just within a few Mpcs. 
  
  Therefore, D, He, Be, and Li  UHECR reach us in  our Local Sheet volume;
see Figure~4. In such a local and inhomogeneous Universe, one should expect quite strong anisotropies. Otherwise, such  strong anisotropy should be unexpected in any hundred Mpc, nearly homogeneous cosmic volume.

\begin{figure}[H]
  \includegraphics[width=0.85\textwidth]{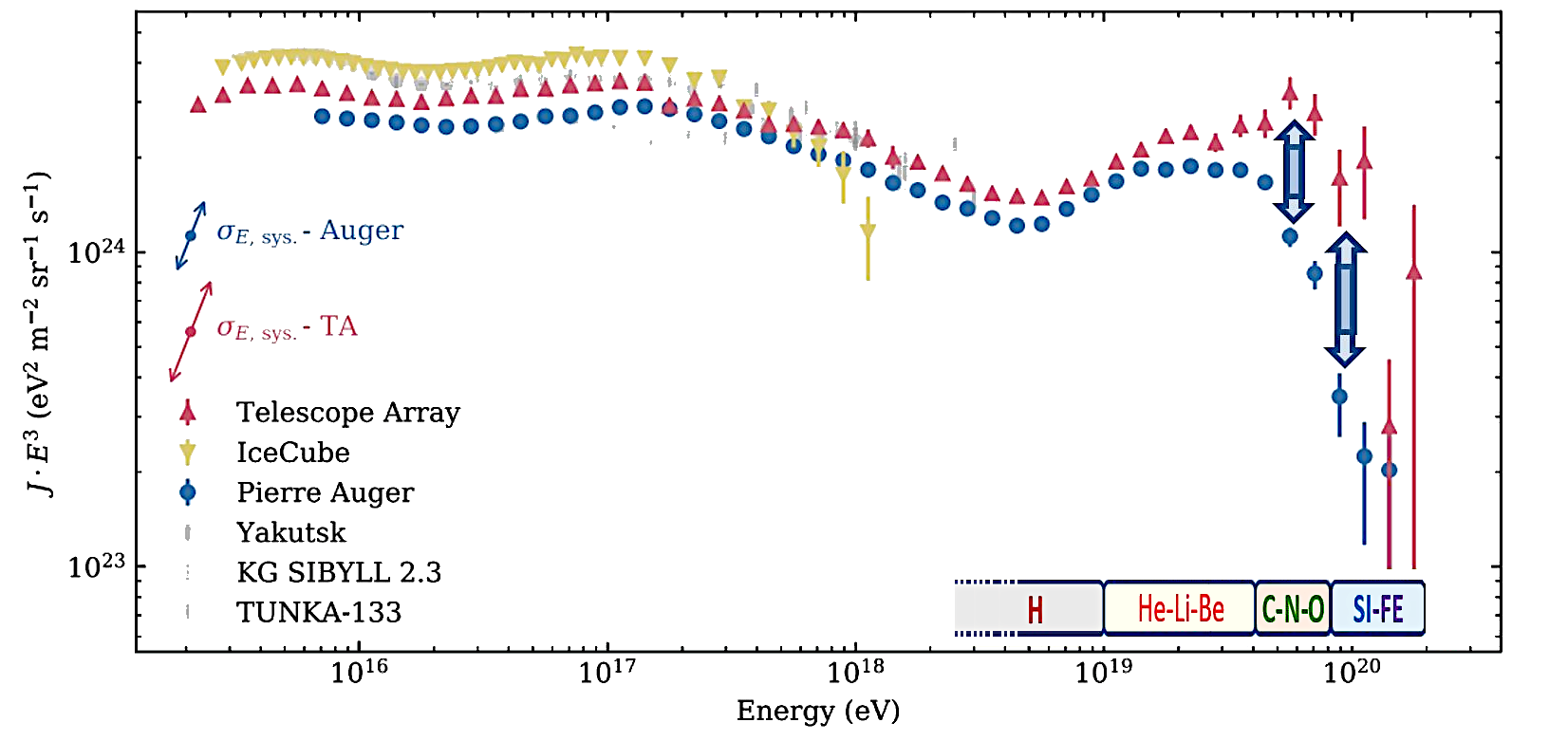}
  \caption[]{
    The latest combined CR spectra from the Auger and TA array detectors. The data are in full agreement up to the energy of several EeV. However, they are not in agreement at higher energies. Northern sky spectra at these edges are harder. Being harder, if they are the lightest nuclei, they might be more directional, keeping a memory of their directionality. 
However, if they are very heavy nuclei with large charges and the highest energy, they should be more bent, leading again to a more smeared map that cancels any mass distribution asymmetry. %
Such asymmetry among the northern and southern sky cannot be explained by any cosmic asymmetry: it needs very few local and nearby asymmetries. 
Indeed, the mass density distribution within the Local Sheet is asymmetric: there is several times more abundant mass density in the north sky than in the south one related to the Local Sheet sources, see Figure $6$ in~\cite{Biteau:2021pru}. 
This naturally explains the puzzling split between the north--south UHECR spectra, as long as the lightest nuclei are the ruling courier.
  At the bottom right, the most probable composition intervals from~\cite{AbdulHalim:20232A} are highlighted and expanded to include the lightest nuclei ($D, He, Li, Be$): the first ones from about 2.5 to 10 EeV UHECR are mostly protons, 
  then the lightest nuclei prevail, followed at higher energies by increasingly heavier nuclei. See Figure~7 and text for further details. \\
\label{UHECRs_EDGES}
}  
\end{figure}

UHECR array detectors exist in both hemispheres. 
Here, we  focus on Telescope Array (TA) data for the Northern Hemisphere and on Auger for the Southern Hemisphere. 
The latter, Auger, covers about four times the surface covered by the former.
Their spectra, after nearly two decades of recording and optimized processing, overlap extensively with a rich sample of data in the energy range $10^{17}$--$10^{19}$ eV. 
However, the two spectra differ remarkably,
only in a narrow window at the highest $E> 3\times 10^{19}$ eV energies~\cite{COLEMAN2023102819},
 as it is shown in Figure~\ref{UHECRs_EDGES}, while seeming to converge again at the highest energy (possibly by \mbox{heaviest nuclei)}.

\unskip

The north--south puzzle discrepancy cannot be reasonably or easily attributed to any experimental error nor to any privilege of Earth observer; it must rather be related to some asymmetry in the mass distribution (or in some sources). As can be assessed from Figure~\ref{UHECRs_EDGES}, the anisotropy is not negligible: it exceeds a factor of $200\%$ near the maximum energy $E = 10^{20}$ eV, well above any experimental error bars or statistical fluctuations.

Such a source of asymmetry should require a huge  disproportion in the mass distribution between the northern and southern sky. Indeed there is several times more abundant mass density in the north sky than in the south one inside our Local Sheet sky; see Figure~$6$ in~\cite{Biteau:2021pru}. 
Therefore, a few active UHECR sources in the northern sky, such as our Andromeda, M31, M82, and Maffei galaxies, may prevail over the high-energy southern ones. We will label in detail some possible candidates in the next galactic and celestial maps.

 The Virgo cluster, the most relevant expected (and absent) cosmic UHECR source for proton composition, is located between the northern and southern equatorial skies, in an unfavorable but anyway an observable location. 
 Its negligible presence is consequential 
 for a lightest nuclei model; as it is filtered and screened, it  results in near opacity. 

\section{The Virgo Absence in UHECR Maps and  Hot Spots}
\begin{figure}[H]
\includegraphics[width = 0.85\textwidth]{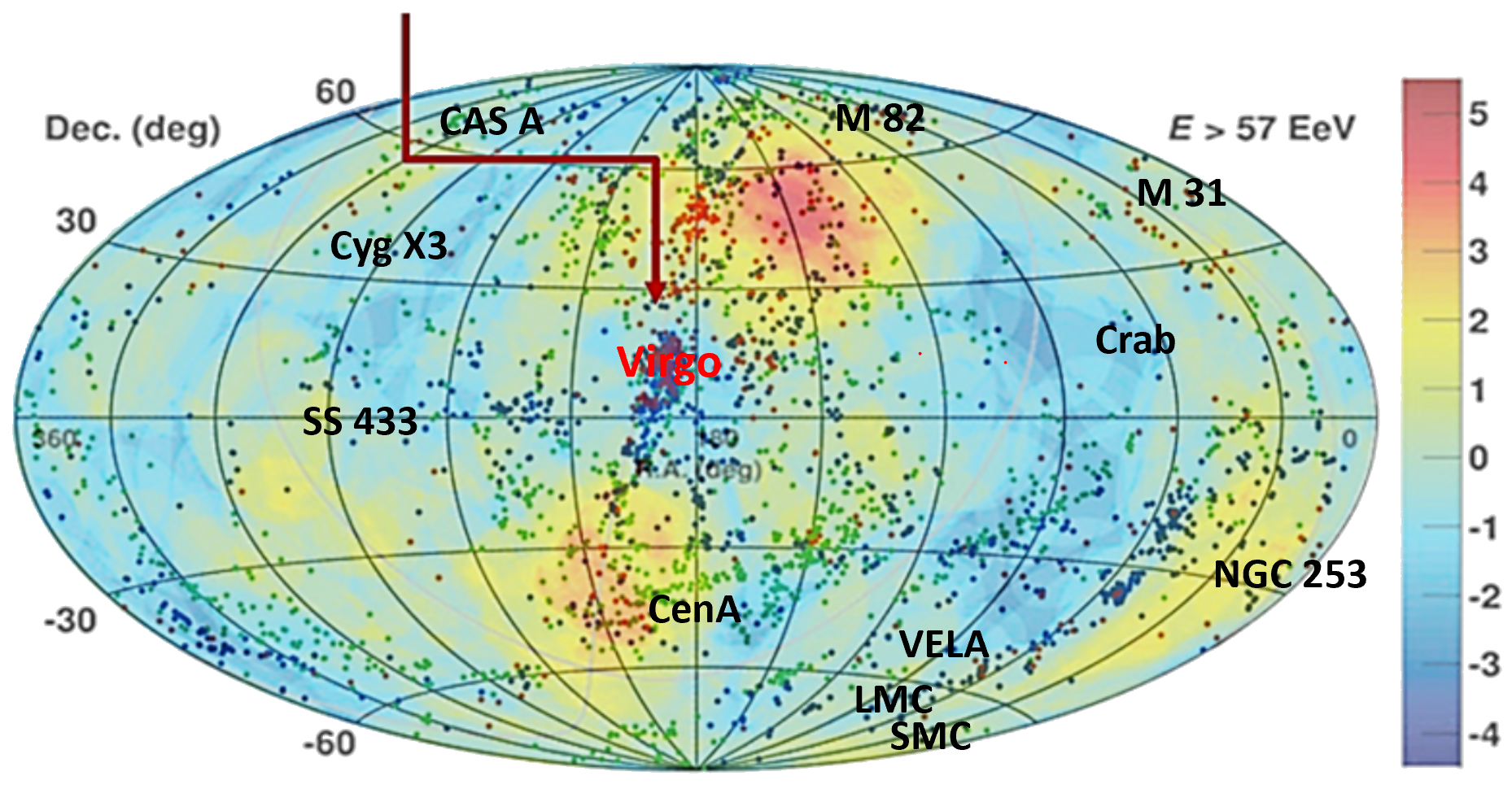}
\caption[]{  
The 
 earliest Auger cosmic map in Hammer celestial coordinates. Dots are the 2MASS galaxies inside the 3000~km~s$^{-1}$ with the heliocentric velocities color-coded with red, blue, and green for increasing red-shift distance~\cite{huchra20122mass}.
The expected UHECR clustering toward Virgo did not emerge as  being the rich and nearest cluster of galaxies at $20$ Mpc. 
This absence had been the main hint leading to the \textit{lightest nuclei UHECR model}~\cite{Fargion:2008sp}. 
On the map, some positions are marked: at the center is the Virgo cluster and its absence.  
A few of the nearest star burst galaxies within 2--3 Mpc are M 82, NGC~253, Cas A, and M31. %
The Cas~A1 is an irregular star burst galaxy located on the prospective in our  galactic plane, where supernovae remnant SNR stand. 
Other galactic sources are the most remarkable SNR or the BH jet ones: Vela, %
 SS433, Cygnus X3, Crab. Other sources may be present in  SMC and LMC (small and large Magellanic cloud)~\cite{FargionICRC2023PoS395}.
 All the marked ones may be possible relics of early SN or GRB-SGRB sources of UHECRs at several EeVs feeding the Auger dipole anisotropy, as shown in the text.
\label{VIRGOABSENCE4}
}
\end{figure}

 The presence of the Virgo cluster, with its thousands of the largest massive galaxies, dominates  the  cosmic GZK volume and the infrared galaxy sky.
 It has been the ideal place to look for UHECR, if they were all or mostly protons. 
 The Virgo cluster sits at the celestial boundary of both the southern and northern sky. Neither detector (TA, Auger) found any such  expected hot spots; see Figure~\ref{VIRGOABSENCE4}. This absence led, as we mentioned, to the earliest proposal of the  \textit{lightest UHECR nuclei with rare heavier nuclei}  model~\cite{Fargion:2008sp,Fargion:2023yiy}.
 
The direct consequence, often neglected or overlooked,
is that UHECR above ten EeV are not made by protons. 
Indeed, thanks to the insight of the pioneering leader of Auger~\cite{Watson:2023uko}, this has been proven to be the case. 
The discovery is based on the air shower slant depth signature.

\begin{figure}[H]
\includegraphics[width = 0.85\textwidth]{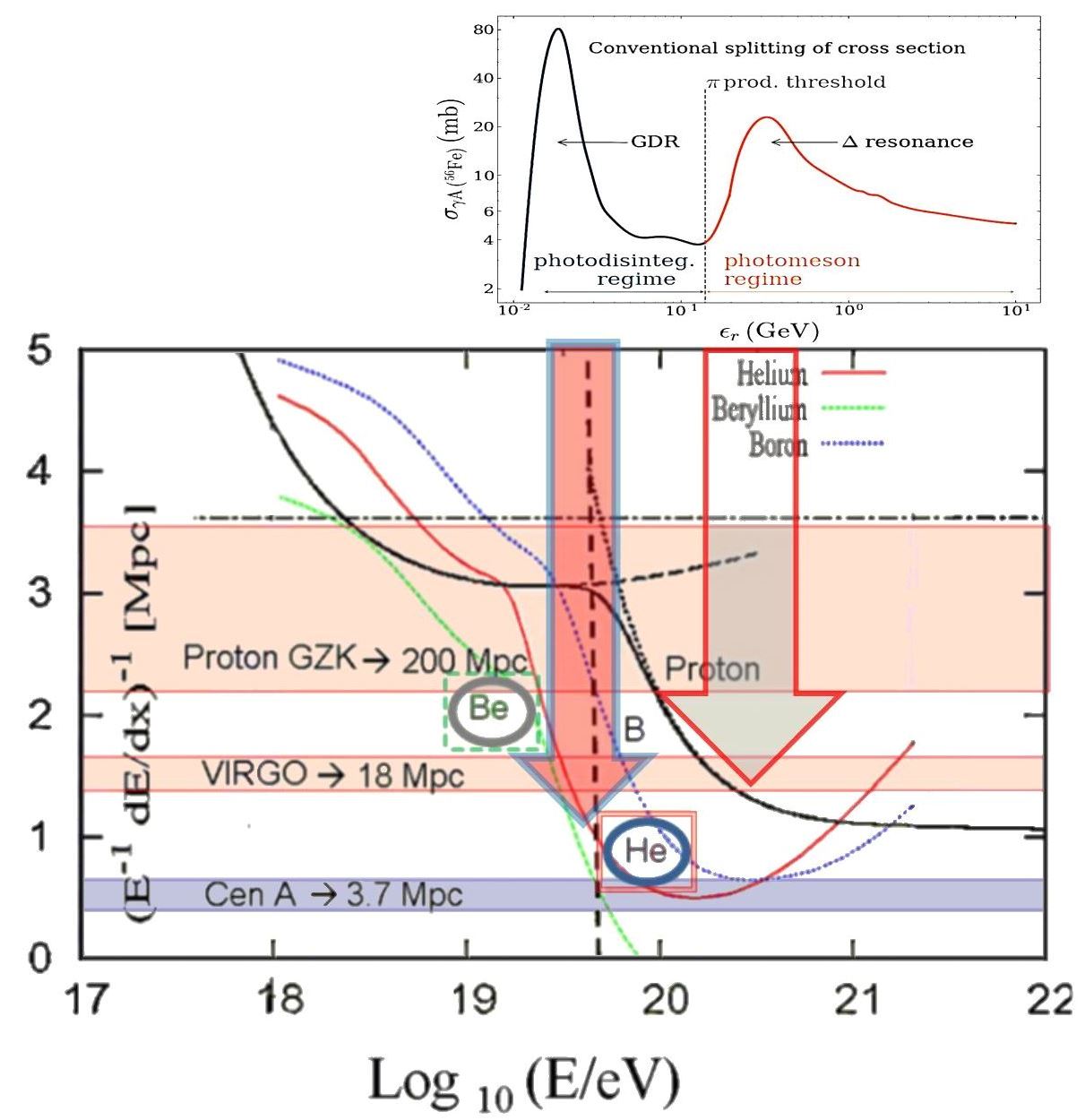}
\caption[]{The allowed UHECR propagation distance, in logarithmic scale, for different energies, for the lightest nuclei and proton. Protons do not suffer from any photonuclear disruption, only the photopion Delta resonance, leading to the longer GZK cut-off. The light, and mainly the  lightest nuclei, suffer from a photonuclear opacity, which are able to be screen signals from Virgo. 
The lightest nuclei also suffer from partial fragmentation, even from a near-4-Mpc distance. Such fragmentation from Cen~A was foreseen and observed  by  their correlated multiplets~\cite{FARGION2011111}.
 See Figure~\ref{Fig5_v05_FINAL}. The photonuclear disruption and the photopion delta resonance roles are shown in an upper panel~\cite{Morejon:2019pfu}. Both the interactions are active in each energy window where UHECRs suffer their characteristic distance cut-off. \label{PHOTONUCLEAR}
}
\end{figure}

Virgo's absence left us so puzzled that, already in $2007$, it forced us to imagine a peculiar \textit{conspirative} phenomenon due to nuclear interactions.
 Is it a filter or an obstacle? 
 The lightest nuclei we noted~\cite{Fargion:2008sp}  should not have been able to survive while arriving from Virgo; see Figure~\ref{PHOTONUCLEAR}.
 But it could reach us from the nearer Cen~A, where the first UHECR clustering was indeed emerging in the Auger map.
As we have noted, there is in fact a much more serious photo-nuclear destruction opacity that occurs mainly only for the lightest nuclei (D, He, Li, Be). They cannot survive  distances as large as the Virgo ones. This was our first assumption to clarify the otherwise unexplained absence of the Virgo cluster from data. 
The presence and dominance of the lightest nuclei in UHECR in early 2008 were quite unexpected and ignored.

     \begin{figure}[H]
\includegraphics[width = 0.85\textwidth]{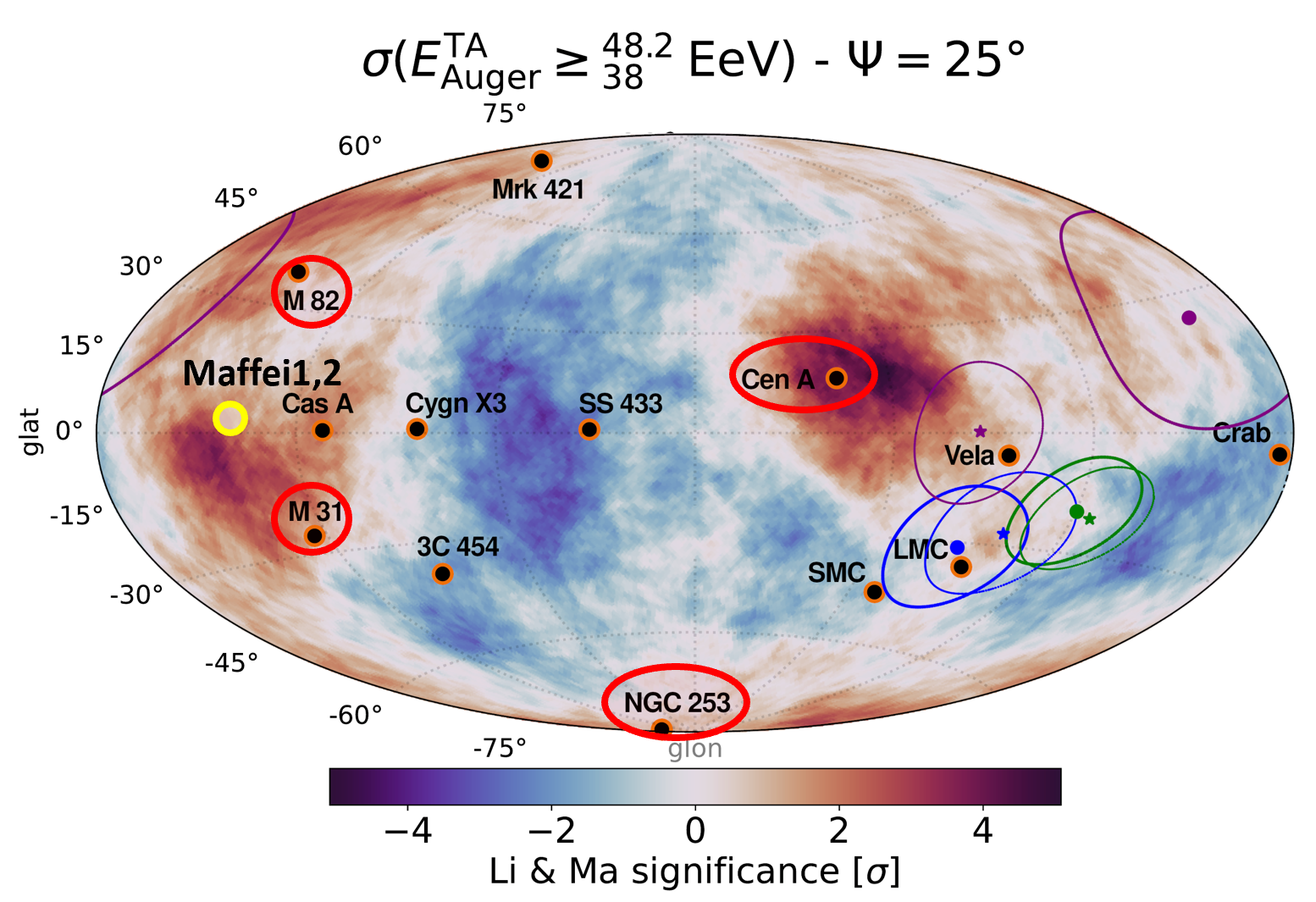} %
\caption[]{
Dipole UHECR anisotropy in galactic coordinates at ten EeV with the overlapping of higher energy 30--50 EeV hot spot clustering. 
In the background is the UHECR clustering statistical weight~\cite{Fujii:2024sys}; see also Figure~\ref{TA_UHECR-2024_FINAL}.
The dipole directions and relative significance are plotted in different colors for different energy ranges: blue, green, and violet, respectively, 8.55--16 EeV, 16--32 EeV, and >32 EeV for Auger and 10--19.4 EeV, 19.4--40.2  EeV,  and >40.2  EeV for TA. Stars and thinner contours represent Auger results obtained by assuming that the moments higher than the dipole are zero \cite{caccianiga2023update}.
Note the proximity of the direction of the most energetic dipole, $E>32$ EeV, to the Cen~A hot spot for $E >38$~EeV.
The NGC~253, a near star burst galaxy, is active  in the bottom edge of the galactic map; it is probably polluting both the new UHECR hot spot energy, and it is feeding the lower energy EeV ones as the Auger dipole. Moreover, at a few or ten EeV, the nearest Vela PSR, the SMC, and LMC sources could all play a role in feeding, with additional multiplets signals,  the same Auger dipole; see Figures~9 and  10.\label{Fig5_v05_FINAL}
}
\end{figure}  
  \unskip   
\begin{figure}[H]
\includegraphics[width = 0.85\textwidth]{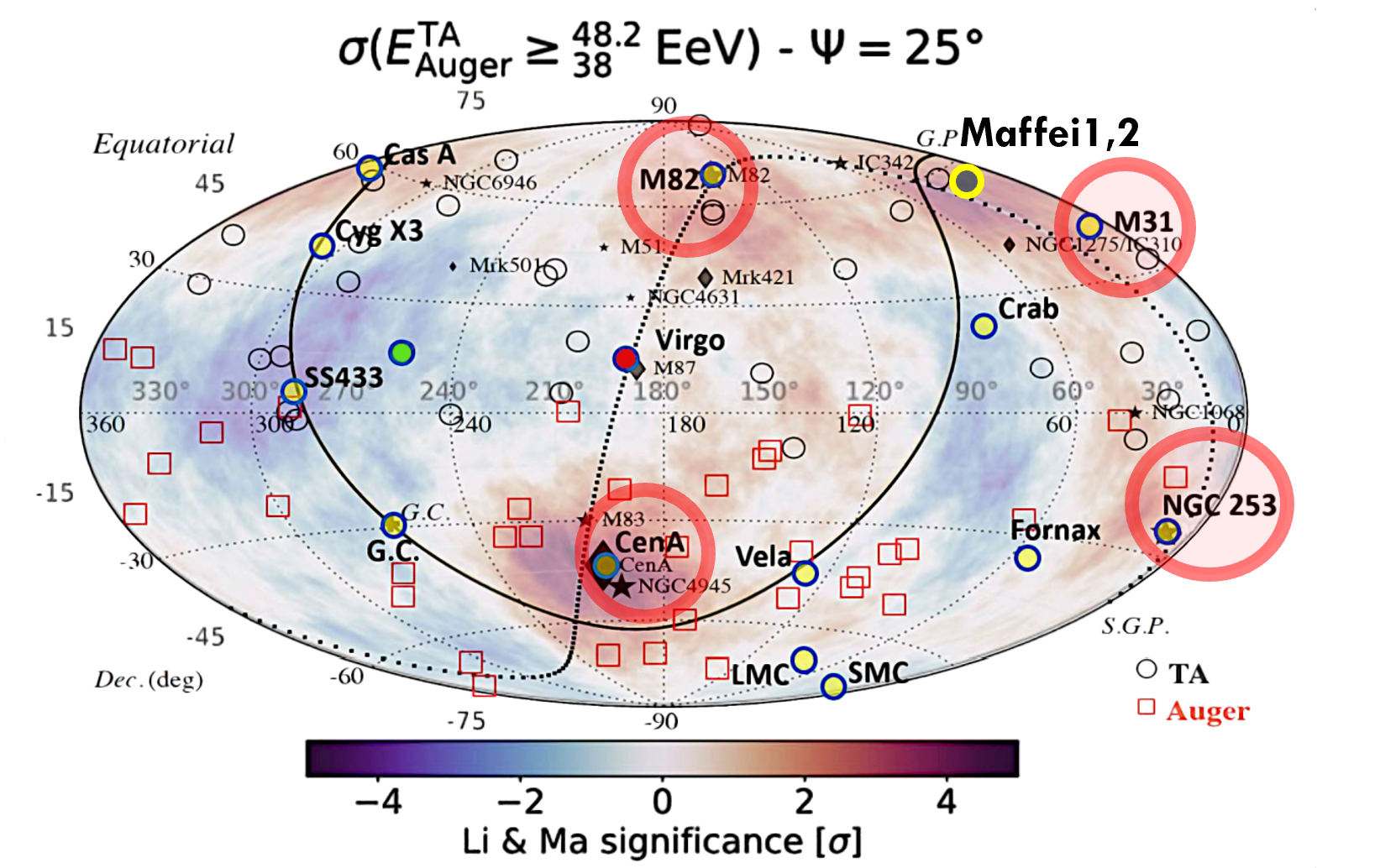}
\caption[]{
\textls[-10]{{{The}  arrival directions of the most recently reported UHECR above 100 EeV measured by Auger and TA, together with nearby astronomical source candidates in Hammer Celestial coordinates. 
In the background, the flux significance sky map of $25^{\circ}$ oversampling for the overlapping region with $E^\textrm{TA}>48.2$~EeV and $E^{\textrm{Auger}}>38$~EeV~\cite{Fujii:2024sys}.
The most energetic event~\cite{TelescopeArray:2023sbd} is tagged with a green  color. Virgo's absence is tagged with a red color.    
The nearest star burst galaxies within  2--3~Mpc (M 82, NGC~253,  and Cas A, as well as M31, Andromeda)  in a local group are considered. The Fornax galaxy,}}}
\label{TA_UHECR-2024_FINAL}
\end{figure}   

{\captionof*{figure}{ not a farther distant cluster, is also considered.
Other galactic sources are shown, such as the most notable SNR or BH jet: Vela, Cas A, SS433, Cygnus X3, and Crab. Other sources may be present in SMC and LMC. Note the four UHECR events overlapping the famous binary jet BH SS433, as well as a doublet occurring in its immediate vicinity, not far from the larger green TA event. 
 Note also a few  UHECR along Cygnus X-3 and Vela.}}
\vspace{12pt}

\begin{figure}[H]

\includegraphics[width = 0.95\textwidth]{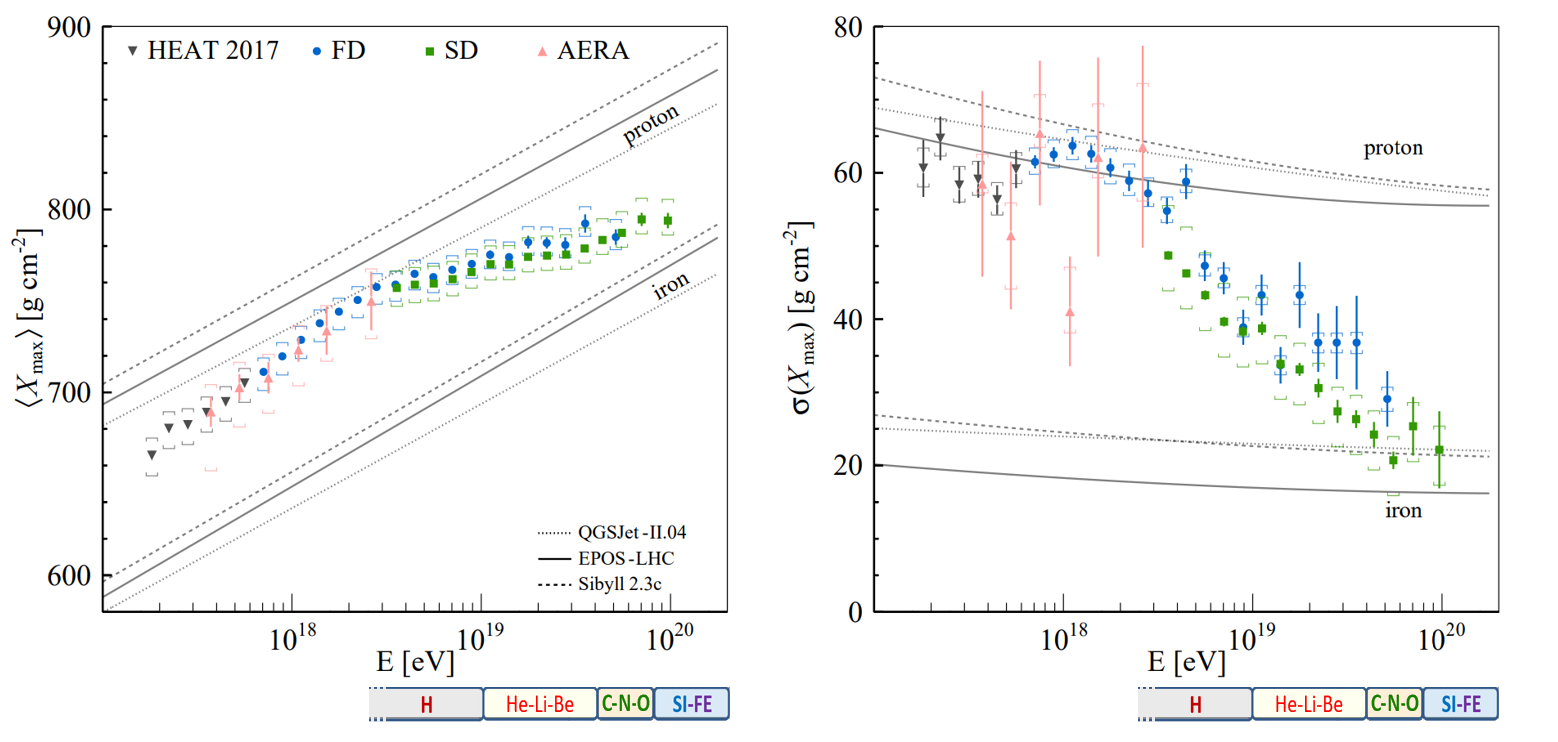}
\caption[]{
The observed UHECR slant depth $X_\textrm{max}$ vs energy on the left panel and  its second moment $\sigma (X_\textrm{max})$ on the right one. At the bottom right of each panel, the most probable composition intervals from~\cite{AbdulHalim:20232A} are highlighted and expanded to include the lightest nuclei ($D, He, Li, Be$). Increasing the energy changes the prevalent carrier. First, protons up to around 10 EeV, then the lightest nuclei in the window 10--50 EeV. Later, light nuclei around 50--70  EeV, and finally, the heaviest ones for 80 EeV up to $2\times 10^{20}$ eV and over.
\label{COMPOSITIONS}
}

\end{figure}

However, soon after these early steps, it was more supported by the  UHECR composition tendency in the Auger data, defended mainly by Alan Watson  at the early beginning~\cite{aab2017combined} and up to recent days~\cite{Watson:2023uko}; see Figure~\ref{COMPOSITIONS}.

 The air shower profile of most UHECRs above ten EeV favors the lightest nuclei, and only at higher energy, light, and heavy nuclei.
The lightest nuclei,  as we  show later (see Figure~\ref{PHOTONUCLEAR}) are filtered from  long distances above a few Mpc, forcing UHECRs  to be located  only within the nearest Local Group, or Local Sheet galaxies. Virgo cannot shine  its UHECR  lightest nuclei to us because of its fragility; see Figure~\ref{VIRGOABSENCE4}.  Cen~A, whose UHECRs  clustering was already arising in early Auger $2007$ data  had the advantage of being nearer  and able to reach us. %
 Cen~A's distance (nearly $4$ Mpc)  could also be able to partially disintegrate the lightest nuclei into smaller fragments. Therefore, it has been expected that about half the energy fragments could be present at a wider bending angle. A possible twin multiplet secondary signature was imagined around Cen~A~\cite{FARGION2011111}. Indeed, these signals were recently observed~\cite{abreu2012search} and, generally, up to now, had gone  unnoticed~\cite{FargionICRC2023PoS395}.

\section{UHECRs: Hot Spots and the Extreme Auger Dipole Asymmetry}

Indeed, the most celebrated Friedmann Universe is both homogeneous and isotropic. Our reference system motion induces a kinetic Doppler asymmetry.

To be more precise,  $2.75$~K cosmic Black Body Radiation has shown tiny anisotropy,  but at the $2\times 10^{-3}$ level. This effect is considerably smaller than the observed asymmetry in UHECR at the Auger dipole sky. It is a kinematic Doppler effect due to our galaxy's motion in the sky, discovered nearly half a century ago. 
The Auger dipole asymmetry at the UHECR edges is $30$ times larger than the cosmic kinetic one:  such a relativistic solution cannot be found.

  In an analogy, Auger discovered a remarkable dipole anisotropy whose amplitude is~$>6\%$: %
   it is not pointing to any cosmic dipole or any far (within GZK) volume cluster; see Figure~\ref{Fig5_v05_FINAL}.
    Once again, this prominent anisotropy is not associated with any cosmic asymmetry, nor with any nearby extragalactic asymmetry, with the exception of Star Burst NGC~253, which in fact is considered by us to be the main candidate source for a hot spot at tens of EeV.
     Again, all this dipole data favor very local, even partially galactic source contributions. 
      In fact, Vela, SMC, and LMC could play an additional role by polluting the anisotropy of UHECR at EeV, where the dipole stands. 

We remind readers that this kind of dipole anisotropy has been  recently~\cite{Kashlinsky_2024} discovered at a few tens of GeV in the Fermi gamma data.
Furthermore, in the last two decades, Milagro, Argo, and HAWC found some lower anisotropies at $10^{-4}$  in the TeV energy windows,  somehow correlated with the same Auger dipole.
Some understanding of such dipole overlapping was attempted as early as 2012~\cite{FargionICRC2023PoS395}.


 \section{Photonuclear Survival Distances for Lightest UHECR Nuclei} 

As we mentioned, the  absence of Virgo is connected to the photo-nuclear disruption of the lightest nuclei.
The photo-nuclear threshold is an order of magnitude lower than the GZK cut-off for the proton and the photopion opacity.
     This difference is due to the role of giant dipole resonance (GDR) in photo-nuclear reactions occurring at around 10--30 MeV compared with the ten times higher %
Delta resonance energy of around $140$ MeV.

Atoms such as deuterium, helium, lithium, and beryllium and their isotopes are so fragile they are constrained to come from a volume of a few Mpc before decaying in flight: 
a Universe within our Local Group or better Local Sheet galaxies.   
This earlier cut-off, the photonuclear disruption,  had been proposed  soon after Auger discovered a possible clustering to Cen~A and  the absence of Virgo~\cite{Fargion:2008sp}: the following TA clustering  toward M82 and the first appearance of the NGC~253 clustering in 2015. Ref. 
\cite{FargionICRC2023PoS395} also confirmed the absence of any clustering toward Virgo.  The lightest UHECR nuclei were confirmed later by Auger composition records and models~\cite{aab2017combined}. 

\subsection*{UHECR 
 Slant Depth, Clustering, and Composition Versus Their Cut-Off} 
     The downward UHECR air shower develops in a tree-like profile, and its  growth and decay shape and corresponding event slant depth  define  the characteristic  nuclear cross-section and most probable  nuclear nature. 
     With the same energy, heavier nuclei interact first at higher altitudes, and lighter ones are deeper where the atmosphere is denser. 
     Slant depths observed by Auger over the last two decades show an evolution with increasing UHECR energy, from a proton-like composition at EeV toward the lightest nuclei, as considered in~\cite{Fargion:2008sp}, and finally to heavy ones at the highest energies; see Figure~\ref{COMPOSITIONS}.
     
The main UHECR capable of offering significant clustering, anisotropy, or hot spots occur above tens or a few tens of EeV. According to experimental data, at energies above these thresholds, the slant depth  is no longer described by the proton but better by the lightest nuclei, then by the light ones, and later by the heaviest ones. The slant depth values show little dispersion, which means little mixing in the UHECR composition.
     
   Starting from lower energies, at about 20--60 EeV, abundant UHECR events occurred. There are enough data to disentangle an anisotropy or a few hot spot clustering. Signals are mainly ruled by the lightest nuclei. Above 50--70 EeV, light nuclei with more charges and deflections should rise, possibly smeared again in an isotropic noise.
   
Finally, at 80--200 EeV, rare and heavier nuclei are more present, with higher energy, higher nuclear charge, and consequent larger bending curvature angle.
      
     The advantage of the lightest nuclei to describe some clustering  led us to consider  a small local Universe, the Local Sheet galaxies, which are less in number and  easier to disentangle. 

    The \textit{UHECR lightest nuclei model}~\cite{Fargion:2008sp}, updated  with  more  details in recent years in~\cite{FARGION2011111} and~\cite{FargionICRC2023PoS395,Fargion:2023yiy}, has remained basically in the same frame since $2008$, as well as in the present article.  
    In this model, the Virgo absence was well explained or even necessary. 
    Cen~A, being the first clustered source, was also tuned to expectations. The subsequent discovery of an additional nearby AGN, M82, and later, Star Burst NGC~253, fit neatly into the model.  
     Cen~A is not close enough to avoid some photo-nuclear destruction. Indeed, we soon predicted the presence of fragments around the Cen~A sources at around twenty EeV. These tracks were soon observed. 
     We had not foreseen similar traces around NGC 253 because its clustering in 2009--2011 was not yet discovered. But these traces are also present; see Figure~10. 
    However, some possible uncorrelated events, neither too near nor too far, may occur above the GZK edges.

   Historically,  in $1991$, at Fly's Eye array detector, the first huge UHECR occurred at $3\times 10^{20}$ eV  energy~\cite{1995ApJ...441..144B}. There was no reasonable UHECR candidate source within a GZK volume for such an event.

   In $1997$~\cite{fargion1999ultra}, a new, alternative neutrino model was offered for such unexplained events. This model is able to correlate far cosmic sources (AGN), well above GZK volume, with signals obtained by their ZeV, (Zevatron or $10^{21}$ eV) incoming neutrino, while scattering onto relic neutrino with tiny masses, at rest, in dark galactic halos. This model, often  called  Z boson resonance or Z burst model~\cite{Weiler:1997sh}, overcomes the GZK cut-off. It is still an allowed model for any unexplained rare cases, as we remind readers in the next section.

Let us now reconsider in more detail the UHECR nuclei deflections.

  Cosmic rays,  being charged and  bent by galactic and cosmic magnetic fields, are flying under coherent or incoherent random deflections.
  Since we observed hundreds of billions of CR records at TeVs-PeVs, their statistical map offered some rare and tiny anisotropies, as little as $10^{-5}$.
   The highest CR,  the UHECR  above several tens of EeV, were expected to lead  to some  multiplet of events. Indeed, such clustering of signals, within  tens of degrees spots  or lower energetic wider dipoles, was finally discovered by Auger, and  by TA,  almost $15$ years ago. 
   
   The detailed opening angle of Lorentz forces is linked to the UHECR nuclei composition, its energy, the magnetic field strengths, and their geometrical structure along the same trajectory.

\section{The Coherent or Random Bending from Local Sheet  Galaxies and Time Delay} \label{7}

 The trajectory of UHECR is deflected by the magnetic fields it encounters between the source and the observer. To calculate the total deflection angle $\alpha_{\mathrm{rm}}$,  because of a random or incoherent walk, it is necessary to take into account various physical characteristics.  More specifically, its nuclear charge number $Z$ and 
 the UHECR energy $E$, as well as its total traveled distance $D$ in an extra-galactic magnetic field $B$. These fields have  a characteristic coherence length path, $d_c$ , and a total distance path $D$.
    
In this view, let us first consider  the UHECR galactic random bending
for a proton $p$, $Z_p=1$, assuming  a total galactic distance  $D =20$~kpc  and a characteristic coherent bending step~\cite{FARGION2011111} 
 of $d_c$ $=1\cdot$~kpc. Therefore, the incoherent deflection angle, in analogy to the  estimate above for the Moon or Sun's shadows,  becomes 
    
\begin{equation}\label{angle}
\alpha^p_{\mathrm{G_{rm}}}=5.65^\circ\left(\frac{Z}{Z_p}\right)\left(\frac{E}{6\times 10^{19}\,\mathrm{eV}}\right)^{-1}\left(\frac{D}{20\,\mathrm{kpc}}\right)^{1/2}\left(\frac{d_c}{\mathrm{kpc}}\right)^{1/2}\left(\frac{B}{3 \mu G}\right).
\end{equation}

The angular separation between observed UHECR smeared anisotropy hot spot and Cen~A is about $18^\circ$.
Assuming for Cen~A a distance of $4$~Mpc and an average magnetic field $B$ of about $3$~nG for the extragalactic path, the angle as a function of energy can be \mbox{rewritten as}

\begin{equation}\label{angle2}
\alpha^p_{\mathrm{E_{rm}}}=3.5^\circ\left(\frac{Z}{Z_p}\right)\left(\frac{E}{6\times 10^{19}\,\mathrm{eV}}\right)^{-1}\left(\frac{D}{4\,\mathrm{Mpc}}\right)^{1/2}\left(\frac{d_c}{\mathrm{Mpc}}\right)^{1/2}\left(\frac{B}{1\,\mathrm{nG}}\right).
\end{equation}

The total bending  angle for a proton is, therefore, the sum of both the angle galactic and extragalactic components,  or $= 9.15^\circ $. 
Indeed, the Helium bending will occur just twice and be as large as the observed $18^\circ$ hot spot along Cen~A, quite close to the observed one. 

 The possibility that fragments of such lightest nuclei also arrive with a correlated tail of events has been foreseen~\cite{FARGION2011111} and soon after observed~\cite{abreu2012search}.  
The multiplet discoveries of two trains of lower energy signals at around 10--20 EeV, both pointing towards Cen~A, are already statistically significant;  indeed, note
 the $+$ tags in Figure~10. 
The probability of hitting such a narrow sky target \textit{a priori} was very low $< 10^{-4}$. 
The more recent evidence of an additional correlation to  the third multiplet was the NGC~253  (unexpected but still observed) pointing to the second hot spot in the Auger sky: NGC~253~\cite{FargionICRC2023PoS395}; see Figure~10. 
The more recent multiples (see Figure~9) found by Auger correlate (by first  approximation) with the Cen~A hot spot, and only partially with the Vela SNR  (or LMC) sources. 
We consider both of these two sources able to contribute to the Auger dipole anisotropy. See Figure~10.  

\begin{figure}[H]
\includegraphics[width = 0.8\textwidth]{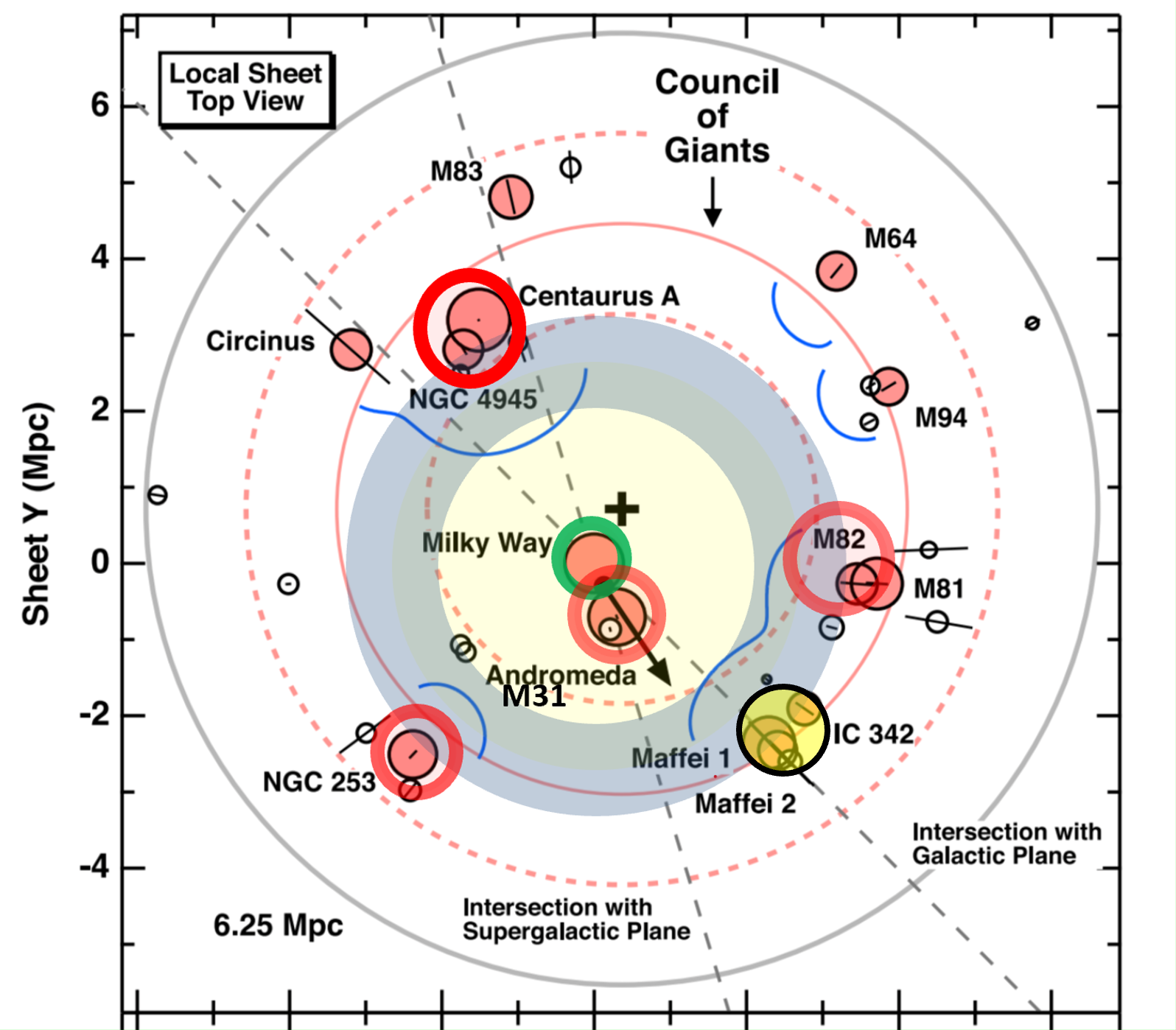}%
\vspace{6pt}

\caption[]{The shadow circle shows the 2--3 Mpc distance from our galaxy. These are the characteristic distances for the UHECR lightest nuclei allowed flight.  The Council of Giants map with a label made by red rings shows the suggested  UHECR candidature for UHECR in the last decade~\cite{FargionICRC2023PoS395}; in $2008$, Auger found a first clustering along Cen~A; later on, TA noted a clustering toward M82, the second hot spot. Later, we noted the possible source of clustering along NGC 253~\cite{FargionICRC2023PoS395}, as well as  a possible role of clustering toward M31 ( Andromeda). We were not aware of the Council of Giants' additional source, namely the Maffei Galaxy 1,2, a missing candidate shown with a yellow ring. 
\label{FOUR+MAP}
}
\end{figure}

  The UHECR photonuclear disruption  survival distance is constrained from the Virgo signals, as shown in Figure~\ref{PHOTONUCLEAR}. However, the Cen~A distance is nearly contained, within \mbox{3--4 Mpc},  in the energy edge  4--6 $\times 10 ^{19}$ eV for  $He$ but not totally contained for $Be$ or heavier ones. See Figure~\ref{FOUR+MAP}.
 Therefore, the presence of the harder UHECR  at 4--6 $\times 10 ^{19}$ eV was expected from Cen~A to lead to fragments at half or less of this energy: $\sim$ 2--3 $\times 10 ^{19}$eV. %

These secondaries of UHECR at ten to twenty EeV were able to confirm the hot spot clustering at 40--60 EeV from Cen~A.  
Indeed, the  following years'  data confirmed this expected multiplet presence.

The observed train of multiplet fragments and their  geometry were compatible with the expected ones~\cite{FARGION2011111}, but the main sources,  Cen~A and later NGC~253, have been well observed; see Figure~10.  The pointing crosses of these UHECR multiplets are well correlated with Cen~A  and also with NGC 253. 

The probability for such a correlation of two events to occur by chance inside a disk of a small radius, nearly $7.5^\circ$, is very small. 
Assuming for Auger an observable solid angle of nearly $3 \pi$, the two disk areas have a probability to hit Cen~A  well below $10^{-4}$. The clustering toward NGC 253 was not foreseen in 2009 but was noted later.

Its additional presence toward the second source candidate contributes to validating the lightest nuclei model scenario.
 
 Let us conclude  by mentioning a few of the largest galactic %
 bending for light nuclei $Si$ or heavy ones, such as the iron or $Ni$ nuclei. We consider
 the nominal  $ 5 \cdot$~kpc  distance up to the well-known  microquasar SS433 distance. 
This system is a famous candidate Pevatron (or even, as we suggest, a source of  UHECR) fed by a precessing jet:
 
 \begin{equation}\label{angle3}
\alpha^p_{\mathrm{G_{rm}}}=2.82^\circ\left(\frac{Z}{Z_p}\right)\left(\frac{E}{6\times 10^{19}\,\mathrm{eV}}\right)^{-1}\left(\frac{D}{5\,\mathrm{kpc}}\right)^{1/2}\left(\frac{d_c}{\mathrm{kpc}}\right)^{1/2}\left(\frac{B}{3 \mu G}\right).
\end{equation}
 
 This bending, amplified by the $Si$ and $Ni$ charges, respectively, for $Z_{Si}=14$ and $Z_{Ni}=28$, becomes, respectively,  $\alpha^{Si} =45.5^\circ$ and $\alpha^{Ni} =91^\circ$. 
 These values are remarkable because they allow the SS433 to be,  in principle,   a \textit{nearby} galactic
 candidate for the most powerful event, named \textit{Amaterasu}, recently reported by TA~\cite{TelescopeArray:2023sbd}.
    
  This event was observed at the highest UHECR energy, $ E = 2.44 \times 10^{20}$ eV, nearly four times larger than the energy considered in the expression above. Therefore, one may derive, respectively, the light and heavy nuclei deflection for the \textit{Amaterasu}  energy \mbox{as follows:} 
  
   \begin{equation}\label{angle4}
\alpha^{Si}_{\mathrm{G_{rm}}}=22.7^\circ\left(\frac{Z}{Z_{Si}}\right)\left(\frac{E}{2.4\times 10^{20}\,\mathrm{eV}}\right)^{-1}\left(\frac{D}{5\,\mathrm{kpc}}\right)^{1/2}\left(\frac{d_c}{\mathrm{kpc}}\right)^{1/2}\left(\frac{B}{3 \mu G}\right).
\end{equation}

  \begin{equation}\label{angle5}
\alpha^{Ni}_{\mathrm{G_{rm}}}=45.5^\circ\left(\frac{Z}{Z_{Ni}}\right)\left(\frac{E}{2.4\times 10^{20}\,\mathrm{eV}}\right)^{-1}\left(\frac{D}{5\,\mathrm{kpc}}\right)^{1/2}\left(\frac{d_c}{\mathrm{kpc}}\right)^{1/2}\left(\frac{B}{3 \mu G}\right).
\end{equation}

Or, simply,  $\alpha^{Si} =22.7^\circ$ . $\alpha^{Ni} =45.5^\circ$. 
These angular deflections for the light nuclei $Si$ are marginally consistent with the SS433-Amaterasu angular distance. The heavy nuclei $Ni$ bending is well compatible with the observed SS433-Amaterasu angular distance of nearly 35$^{\circ}$--40$^{\circ}$; see Figure~\ref{TA_UHECR-2024_FINAL}.
 
The same heavy nuclei $Ni$-like UHECR can be coherently deflected by the nearby Vela SNR, $0.29$ kpc, by an angle comparable to the observed displaced clustering. The same train of events might feed the  UHECR EeV dipole clustering in Auger events: $18.7^\circ$. 

Also, LMC and SMC could be sources of this multiplet at a few or ten EeV,  feeding the Auger dipole.

  The consequences of such an Auger dipole are  secondary traces of UHECR that might be correlated  with very recent gamma GeV dipole anisotropy~\cite{Kashlinsky_2024}. 
The gamma rays dipole anisotropy observed at tens of GeV could also be traced and overlapped to the same UHECR signal. 
   It could be worthwhile to verify,  in the future, if there is also such a neutrino anisotropy.
Eventual  consequent  $\tau$ neutrino  signals at PeV should be searched and taken into account~\cite{FARGION2011111,fargion2002discovering,fargion2004tau}.

We note that the nearby Vela SNR, our brightest galactic gamma source, may be capable of ejecting heavy UHECR at wide angles.
At such a close distance, we use the  coherent-bending formula:
 
\begin{equation}\label{angle6}
\alpha^{Ni}_{\mathrm{G_{Coh}}}=23.6^\circ\left(\frac{Z}{Z_{Ni}}\right)\left(\frac{E}{6\times 10^{19}\,\mathrm{eV}}\right)^{-1}\left(\frac{D}{0.29\,\mathrm{kpc}}\right)\left(\frac{B}{3 \mu G}\right)
\end{equation}

\begin{figure}[H]
\includegraphics[width = 0.9\textwidth]{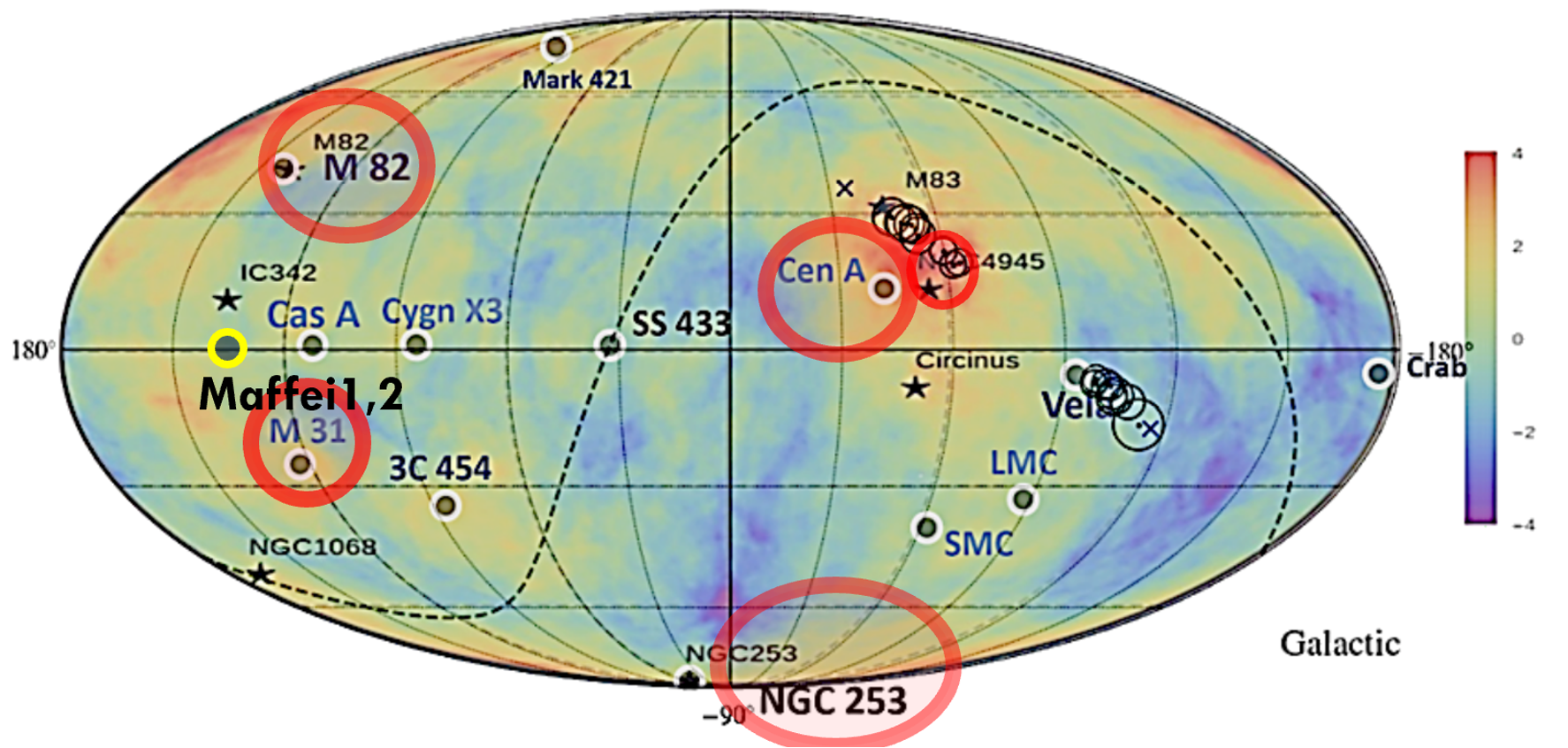}%
\caption[]{The most recent multiplet clustering  in galactic coordinate~\cite{Aab_2020}. Several nearest SNR and star burst galaxies within 2--3~Mpc: Cen~A, M 82, NGC~253, Cas~A. Note the upper multiplet pointing  toward the Cen~A hot spot, possibly to a more distant star burst source, M83, at $5$ Mpc distance. Additional clustering lays along Vela SNR,  but not in a successful overlapping. Indeed, a few UHECR composition differences might invert the direction arrow of the multiplet clustering. The Vela, with the LMC  role, could be important to feed the observed widespread Auger dipole anisotropy. 
These multiplets, again, are shown over  the UHECR map clustering considered by early authors~\cite{ANCHORDOQUI20191}. \label{MULTIPLET_FINAL}
}
\end{figure}

Such UHECR   multiplets are possibly the ones observed; see Figures~\ref{TA_UHECR-2024_FINAL} and \ref{MULTIPLET_FINAL}.

In an analogy, the bending of comparable  heaviest $Ni$ UHECRs ejected from LMC and SMC can feed and pollute the observed Auger dipole anisotropy at several EeVs or at  10--20 EeV. More signals and clustering  of the  UHECR composition maps will prove this possibility. 
Their approximated galactic signature could be  present  in the recent UHECR composition maps~\cite{Mayotte2022Aq}.

Let us now remind readers of the time flight for such a random bending trajectory.

The   flight delay between the emission and arrival of a UHECR proton compared with a direct photon, due to the random walk (for example, from a distant Cosmic 100~$\mathrm{Mpc}$), would be

\begin{adjustwidth}{-\extralength}{0cm}
 \begin{equation}\label{time}
\Delta t\simeq\frac{D\alpha_{\mathrm{rm}^2}}{4c}\sim3.75\times 10^6
\left(\frac{Z}{Z_{p}}\right)^{2}     \left(\frac{E}{6\times 10^{19}\,\mathrm{eV}}\right)^{-2}\left(\frac{D}{100\,\mathrm{Mpc}}\right)^{2}\left(\frac{d_c}{\mathrm{Mpc}}\right)\left(\frac{B}{3\,\mathrm{nG}}\right)^2\,\mathrm{yr}.
\end{equation}
\end{adjustwidth}

  Such a flight delay is so long that there might be no possible ( or probable) correlation among the individual  far, hundred Mpc, gamma AGN activity~\cite{FargionICRC2023PoS395}, and UHECR clustering. Indeed, the AGN lifetime is reasonably a few hundred thousand years, while the consequent delayed arrival for the UHECR flight is much longer. 
  
  Therefore, far cosmic source connection (as Mrk 421, 3C454, or Perseus Cluster) by UHECR nuclei (or proton) random flight (or walk) is not an acceptable option. 
  
  Nevertheless, any far cosmic  ZeV UHE, $10^{21}$ eV, neutrino-hitting relic antineutrino at rest in a Mpc, hot dark matter galactic halo,   being mostly along  a straight flight, will show, after their interaction,  a little time delay and a narrow nucleon clustering. It could  be well timed with the observed photon flare activity~\cite{fargion1999ultra,FargionICRC2023PoS395}, as discussed in the  next section.
    The recent (weak) UHECR clustering along the very bright and distant Markarian 3C $454$ (see Figure~\ref{MULTIPLET_FINAL}) could be such a signal by  such a candidate to  be carefully considered.
    If those UHECR secondaries are nucleon (proton-like) by composition, the Z resonant neutrino model might be the correct one.

\section{Amaterasu: The Revival of a Z Burst Model or a Heavy UHECR Nuclei from SS433?}

  In 2021, $30$ years after Fly's Eye,  an event %
 of comparable energy once again occurred in the North Telescope Array sky. 
 Its discovery has just recently been published~\cite{TelescopeArray:2023sbd}; see Figure~\ref{TA_UHECR-2024_FINAL}.

 Both of these huge northern sky events have no obvious nearby correlated  sources located within their GZK volumes and solid angle view, which is a very surprising and still unexplained signal. 
   In our present lightest nuclei model, there is only room for a few heavier nuclei at the highest energies. The highest  could be  mostly Nickel- or Iron-like nuclei.
The bending by the galactic fields in this case can reach several tens of degrees, and therefore, their original directionality will be confused and lost. %
 
 The presence of several nearby galactic plane doublets and the largest recent event~\cite{TelescopeArray:2023sbd} could  also be  related %
to galactic early sources, such as SS433 or Cygnus X-3,  in their brightest activity at their birth.
 
Other solutions for UHECR uncorrelated sources have been considered in early Fly's  Eye  array discovery. First, an exotic  heavy   unstable relic particle  whose decay could populate the  UHECR sky anywhere. Another  model, which is recalled in detail later, is based on relic neutrinos hit by ZeV neutrinos: their ultrarelativistic $Z$ boson decay in flight could feed the UHECR signals.

This early proposal~\cite{fargion1999ultra} somehow was related  to the  neutrino mass discovery  in early 1980--1997. It was based on the existence of a UHE ZeV neutrino ejected from far sources hitting relic ones (one or few $eV$  mass) on a dark hot halo, interacting by Z boson resonance. 
Such ZeV neutrinos have the role of a silent courier,  able to overcome any GZK opacity,  reaching us from any cosmic edge. The model was also based on the presence of clustered cosmic relic neutrino, requiring a tiny, at best around 0.4 eV mass, wide hot halo. Present neutrino mass bounds favor near or lower masses than 0.4~eV, requiring extreme tens of ZeV neutrino energy~\cite{fargion2001clustering}. 

Anyway, such a dark cloud could be an effective target for ZeV neutrinos, being an ideal beam dump calorimeter. Indeed, the large, peaked neutrino--antineutrino scattering cross-section into Z boson resonance could be able to lead to several final UHECR hadrons via its secondaries~\cite{fargion1999ultra,Weiler:1997sh}. Among them are the UHE nucleons and anti-nucleons that will appear in the terrestrial atmosphere as the observed UHECR at tens or hundreds of EeV~energies.

Incidentally, we  remind readers  that the electromagnetic Z boson secondaries are also decaying in pions and, later on, in TeV-PeV photon energies~\cite{protheroe2000infrared}. This process  may also successfully overcome the  infrared-TeV opacity from far cosmic flight.
This offers a possible solution to the  recent,  puzzling discovery of tens of TeV gamma photons observed during GRB 221009A by LHAASO array~\cite{LHAASO:2023lkv}. 
Indeed, in this Z boson model, the UHE secondary propagation   
could offer the observed  UHECR signals, as well as the TeV-PeV gamma signals, overcoming the infrared-TeV cut-off. 
The  neutrino masses needed for this Z boson model could range from 0.1 to 0.4 eV~\cite{fargion2001clustering}.

Being nearly consistent with present cosmology bounds, they require an exceptional energetic ZeV neutrino signal. Also, a slightly larger value for the neutrino mass, $1.6$~eV, could be  tuned with the  recent sterile neutrino claims. We mention this model (often referred to as the Z burst one~\cite{fargion1999ultra, Weiler:1997sh}) just as a possible
solution for the few unsolved clustering, such  as the recent \textit{Amaterasu}, the most extreme one detected by TA~\cite{TelescopeArray:2023sbd} (see Figure~\ref{TA_UHECR-2024_FINAL}), as well as for the 3C 454 eventual hot spot clustering, both with the Fermi gamma spot and the Auger UHECR mild clustering. 
On the same subject, there are  additional correlations among the UHECR and the active flaring activities. As from Mrk 421 in 2012 and 2013~\cite{Fraija:2023yil}, all such far cosmic connections  might require a Z burst model as their~solution.

Anyway,  a  more conventional solution, as mentioned above, is based on the heaviest  UHECR nuclei  and their largest deflection. The event can have originated,  for instance, from a galactic SS433 source;  it is a well-known galactic binary system whose jet is spinning and precessing in spirals. At its birth, thousands of years ago, it could have been ejecting UHECR energy  signals in a GRB beam, observed with a delay, nowadays in TA. 


\section{Discussion and Conclusions}

CR are charged and bent. At the highest energy,  above EeV, $10^{18}$ eV, the bending for such UHECRs might be  reduced; indeed, at tens of EeV, some final clustering  was aroused in Auger and TA data since 2007; more recently, a  dipole around  $10^{19}$ eV was  found, and also, about three hot spots at $4\times 10^{19}$ eV were noted.

The nuclear cut-off for UHECR, either by  photopion or by photo-disruption (see \mbox{Figure~\ref{PHOTONUCLEAR}}), bound the UHECR in a near GZK cosmos for nucleon (or heavy nuclei), or even into a much smaller one for lightest nuclei (He, D, Li, Be).
 The Auger and the TA  clustering did not show the expected Virgo mass presence within the GZK volume. Therefore, it has been claimed that UHECRs were dominated by the lightest nuclei, such as UHECR  at $4\times 10^{19}$ eV.
The recent composition signature of UHECR indeed showed such a clear evolution:  from proton at a few EeV  to the lightest nuclei at  $4\times 10^{19}$ eV, and a few light-to-heavy (Nickel-like) nuclei composition above $8\times 10^{19}$ eV energy. 
Therefore, we first meet at a few to several EeV of energy, a wide, smeared UHECR clustering, such  as the Auger dipole one. Then, at higher tens of EeV energy,  we encounter more collimated signals  from the lightest nuclei forming the hot spot ones. 
Later on, the appearance of  light nuclei, with larger charges,  offers a smoother, isotropic map again. Finally,  at the largest energy of 80--100 EeV,  signals are due to a more bent UHECR behavior because of the  largest charge in the heaviest nuclei. They might even be contained or originated in our own galaxy and in the Local Sheet galaxies.

In a few sentences, UHECR at EeV energies and below are  mostly galactic spiraling CR  smeared as protons and partially as lightest nuclei. At a few EeV, they  form in a wide asymmetric dipole  made by LMC, Vela, Crab, and NGC 253. Then, at a few tens of EeV, UHECR are clustered in a few hot spots originating from the nearest Local Sheet star burst galaxies, all bounded by photonuclear disruption. Cen~A and NGC 253 rule in the South Auger sky. M82, M31, Cas A, and Maffei galaxies rule in UHECR, TA, and the North sky.  Finally, most energetic UHECR made by the heaviest nuclei  are much more bent and smeared and mainly originate (or are captured) from our magnetic fields by galactic or nearby  Local \mbox{Sheet sources.} 


Over the past two decades, the prevalent view first favored proton carriers and extragalactic cosmic AGN jets in super cluster planes as the ideal accelerator for UHECR.
The apparent GZK cut-off by Hires,  Auger, and TA  also favored this view.  But the Virgo absence, as discussed above, and the lightest or heavy nuclei signature in  UHECR air shower profiles, are changing the models' frames.

Since 2008, the idea that UHECR are mostly the lightest nuclei has been developed, reaching us mainly from the nearest Local Group, or to be more detailed, within the Local Sheet, i.e., from Cen~A, M82, NGC~253, and probably from the irregular nearby Galaxy Cas~A. 
Even our nearby M31, Andromeda, could be the source of a recent growing clustering of UHECR events; see Figure~\ref{Fig5_v05_FINAL}. The  additional role of the Maffei Galaxies, well within the nearest few Mpc distance, has been added in the present article. They are possibly correlated, with M31, with a recent growing clustering of UHECR events in their surrounding space.
    They are all in our Local Sheet Universe.

The presence of multiplets pointing to Cen~A and, in particular, of the last NGC~253~\cite{Fargion:2023yiy}, had been noted to be  feeding  the Auger dipole anisotropy~\cite{Fargion:2023yiy} Figure~\ref{Fig5_v05_FINAL}).  The most recent multiplet is consistent with the Cen~A and Vela or LMC possible roles (see Figure~\ref{MULTIPLET_FINAL}) also possibly feeding the Auger dipole.

Several EeV dipole anisotropies could also be  polluted by a few, well-known,  galactic transient sources, such as the ones in LMC,  SMC, and the small BH jet in star burst NGC~253. 

A remarkable correlation with UHECR quadruplets around SS433 (a famous precessing jet by a microquasar BH) is underlined. 
Eventual heavy nuclei, such as $Ni $ atoms, could be so heavy and charged to be  very bent, nearly $45.5^{\circ}$ from such a nearby $5$ kpc distance. 
It could therefore even point toward and correlate with the recent \textit{Amaterasu} event, the highest UHECRs ever observed by TA~\cite{TelescopeArray:2023sbd}; see  Figure~\ref{TA_UHECR-2024_FINAL}. 

\begin{figure}[H]
 \includegraphics[width = 0.85\textwidth]{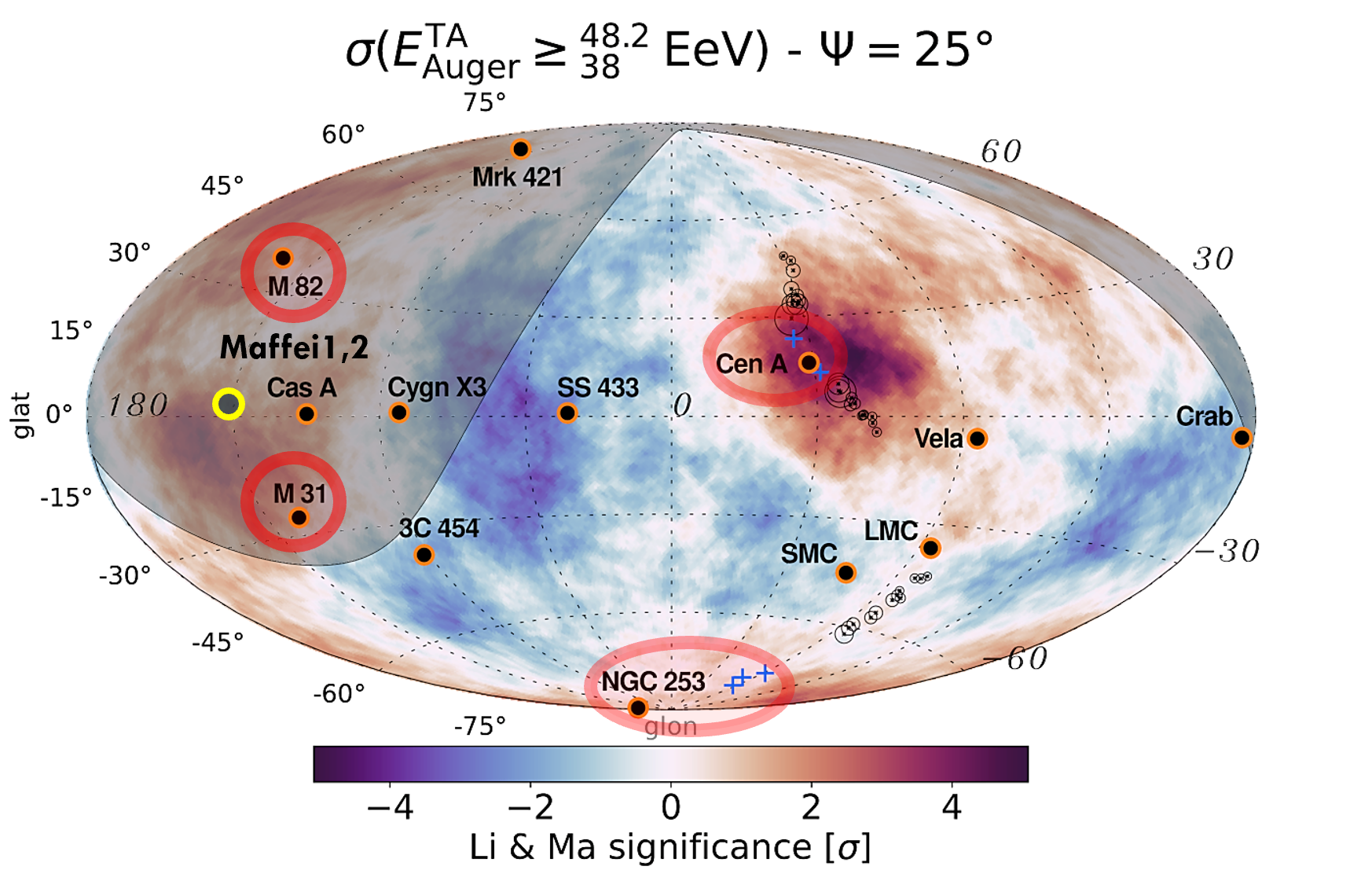}%
\caption[]{The clustering and the main sources with earliest multiplets in galactic coordinate. Several nearest SNR and star burst galaxies within 2--3~Mpc: Cen~A, M 82, NGC~253, Cas A. The last one is a nearby irregular star burst galaxy located on a galactic plane, also in axis with its SNR at the Milky Way plane. The three UHECR multiplets~\cite{abreu2012search,FargionICRC2023PoS395} are estimated to come from the few candidate sources, marked by the blue ``+'',  two toward Cen~A and one toward NGC 253, which are overlapping the most recent clustering map based on TA and Auger data~\cite{Fujii:2024sys}. \label{Fig8_v06_FINAL}
}
\end{figure}

In summary, the lightest nuclei model in a Local Group and partially in nearly galactic sources is suggested.  it is supported by different hints: a recent UHECR asymmetry among northern--southern sky; an anisotropy due to the Local Sheet mass density asymmetry (higher in the North equatorial sky than the South one). This is because of the absence of the Virgo Cluster in UHECR Auger data, due to the lightest nuclei dominance in tens EeV energy sky, as well as 
because of the compositional transition of UHECR from protons at  ten EeV to the lightest nuclei (He, Li, Be) at a few tens of EeV. This is based on how slant depths have been pointing in the last decade, and by the remarkable multiplets that have been somehow foreseen nearby AGN, Cen~A, and NGC~253~\cite{FargionICRC2023PoS395}~(see Figure~\ref{Fig8_v06_FINAL}). 

Other recent models proposed local origins for UHECR in similar frame, 
based mainly on the star burst assumption~\cite{Taylor:2023qdy}. There are overlapping views of these models. 

The tuned and bounded local Universe size for us is basically linked to the  photo disruption of the lightest nuclei; this key role was solving the Virgo absence, 
the \textit{corner-stone} on the road toward a local universe for UHECR.

The absence of any  galactic plane in the UHECR signature, such as the Auger dipole,  suggests  the absence of any continuous spherical averaged emission, such as the SNR galactic plane. It favors a rare, beamed, and accidental source location in our own Galaxy or in \mbox{LMC volume}. 

The unexpected directions of UHECR as the two most intense UHECR in Fly's Eye and in the recent TA event, Amaterasu,  could  probably be the heaviest nuclei, such as Fe and Ni. They are more energetic than others, being at hundreds EeV; but they are at least ten or twenty times more charged and bent  than any lightest nuclei trajectories at tens of EeV. 
    This makes it more acceptable for the lightest nuclei UHECR clustering at a few tens of EeV  with respect to a blurred and smeared hundred EeV nuclei.

UHECRs may rise also from past galactic jet events at peak intensity, such as, for instance from SS433, Cas~A, Cygnus X-3, Vela, LMC, SMC, or Crab. 

The unexplained hundreds of EeV sources, if  from the nucleon,  may be the exotic neutrino \mbox{Z burst model.} 

Indeed, any UHE ZeV  neutrino scattering onto relic ones could still offer an alternative  solution~\cite{fargion1999ultra} based on neutrino mass in dark warm halos and on UHE ZeV cosmic neutrino annihilation via Z boson resonance.

The key to disentangling the two models is the nature of their composition. 
A nucleon nature, from shower slant depth, for the Amaterasu event, suggests a far source bursting via the UHE ZeV neutrino model.
An Amaterasu heavy nuclei composition favors  a very local, maybe galactic source via its heaviest and most bent trajectory. 

The conventional model of the lightest UHECR nuclei from the few Local Sheet sources, such as Cen~A, NGC 253, Andromeda, M82, and Maffei Galaxies, with some contributing from the heaviest ones at the highest energies even from our own galaxy, currently seems the most reasonable and acceptable one to encompass most UHECR puzzles.

\subsection*{Note}

Since this paper was submitted, a few works with similar models have been published simultaneously, e.g.,~Ref. \cite{Marafico:2024qgh}. In particular, there is a recent one on the role of the highest energy heavy nuclei feeding the observed isotropic sky of UHECR~\cite{TelescopeArray:2024oux}, as well as a second more updated one on the preferential map for the clustering of UHECR  along the nearest sources in the Super-galactic plane~\cite{PierreAuger:2024hrj}. 
Our results overlap, at least in part, with the findings of all \mbox{three papers.}

\authorcontributions{Conceptualization, D.F. and M.Y.K.; methodology, D.F.; software, D.F. and P.G.D.S.L.; validation, D.F., P.G.D.S.L. and M.Y.K.;  investigation, D.F.; resources, D.F. and P.G.D.S.L.; writing---original draft preparation, D.F.; writing---review and editing, D.F. and M.Y.K. All authors have read and agreed to the published version of the manuscript.}

\funding{The work by M.K. was performed with the financial support provided by the Russian Ministry of Science and Higher Education, project “Fundamental and applied research of cosmic rays”, No.~FSWU-2023-0068.}

\dataavailability{{The original contributions presented in the study are included in the article. Further inquiries can be directed to the corresponding author.}} 

\conflictsofinterest{The authors declare no conflict of interest. The funders had no role in the design of the study; in the collection, analyses, or interpretation of data; in the writing of the manuscript, or in the decision to publish the~results.} 



\begin{adjustwidth}{-\extralength}{0cm}

\reftitle{References}

\externalbibliography{yes}

\PublishersNote{}
\end{adjustwidth}

\begin{thebibliography}{999}

\bibitem[Fargion(1981{\natexlab{a}})]{Fargion:1981ge}
Fargion, D.
\newblock {Deflection of Massive Neutrinos by Gravitational Fields}.
\newblock {\em Lett. Nuovo Cim.} {\bf 1981}, {\em 31},~49.
\newblock {\url{https://doi.org/10.1007/BF02788167}}.

\bibitem[Fargion(1981{\natexlab{b}})]{Fargion:1981gg}
Fargion, D.
\newblock {Time delay between gravitational waves and neutrino burst from a
  supernovae explosion: A Test for the neutrino mass}.
\newblock {\em Lett. Nuovo Cim.} {\bf 1981}, {\em 31},~499--500.
\newblock {\url{https://doi.org/10.1007/BF02778100}}.

\bibitem[Abbasi et~al.(2022)]{IceCube:2022der}
Abbasi, R.;  Ackermann, M.; Adams, J.; Aguilar, J.A.; Ahlers, M.; Ahrens, M.; Alameddine, J.M.; Alispach, C.;  Alves, A.A., Jr.; Amin, N.M.;  et~al. [IceCube Collaboration].
\newblock {Evidence for neutrino emission from the nearby active galaxy NGC
  1068}.
\newblock {\em Science} {\bf 2022}, {\em 378},~538--543.
\newblock {\url{https://doi.org/10.1126/science.abg3395}}.

\bibitem[Datta et~al.(2005)Datta, Fargion, and Mele]{Datta:2004sr}
Datta, A.; Fargion, D.; Mele, B.
\newblock {SUSY resonances from UHE neutralinos in neutrino telescopes and in
  the sky}.
\newblock {\em JHEP} {\bf 2005}, {\em 9},~007.
\newblock {\url{https://doi.org/10.1088/1126-6708/2005/09/007}}.

\bibitem[Fargion et~al.(2004)Fargion, {De Sanctis Lucentini}, De~Santis, and
  Grossi]{fargion2004tau}
Fargion, D.; {De Sanctis Lucentini}, P.G.; De~Santis, M.; Grossi, M.
\newblock Tau air showers from Earth.
\newblock {\em  Astrophys. J.} {\bf 2004}, {\em 613},~1285.
\newblock {\url{https://doi.org/10.1086/423124}}.

\bibitem[Fargion(1999)]{fargion1999nature}
Fargion, D.
\newblock On the nature of GRB-SGRs blazing jets.
\newblock {\em Astron. Astrophys. Suppl. Ser.} {\bf 1999}, {\em
  138},~507--508.

\bibitem[Grieder(2001)]{Grieder:2001ct}
Grieder, P.K.F.
\newblock {\em {Cosmic Rays at Earth: Researcher's Reference, Manual and Data
  Book}}; Elsevier: Amsterdam, The Netherland,  2001.

\bibitem[Soldin et~al.(2024)Soldin, Evenson, Kolanoski, and
  Watson]{SOLDIN2024102992}
Soldin, D.; Evenson, P.; Kolanoski, H.; Watson, A.
\newblock Cosmic-ray physics at the South Pole.
\newblock {\em Astropart. Phys.} {\bf 2024}, {\em 161},~102992.
\newblock
  {\url{https://doi.org/10.1016/j.astropartphys.2024.102992}}.

\bibitem[Greisen(1966)]{greisen1966end}
Greisen, K.
\newblock End to the cosmic-ray spectrum?
\newblock {\em Phys. Rev. Lett.} {\bf 1966}, {\em 16},~748--750.
\newblock {\url{https://doi.org/10.1103/PhysRevLett.16.748}}.

\bibitem[{Zatsepin} and {Kuz'min}(1966)]{1966JETPL...4...78Z}
{Zatsepin}, G.T.; {Kuz'min}, V.A.
\newblock {Upper Limit of the Spectrum of Cosmic Rays}.
\newblock {\em Sov. J. Exp. Theor. Phys. Lett.}
  {\bf 1966}, {\em 4},~78.

\bibitem[Fargion(2008)]{Fargion:2008sp}
Fargion, D.
\newblock {Light Nuclei solving Auger puzzles?}
\newblock {\em Phys. Scripta} {\bf 2008}, {\em 78},~045901.
\newblock {\url{https://doi.org/10.1088/0031-8949/78/04/045901}}.

\bibitem[McCall(2014)]{McCall:2014eha}
McCall, M.L.
\newblock {A Council of Giants}.
\newblock {\em Mon. Not. Roy. Astron. Soc.} {\bf 2014}, {\em 440},~405--426.
\newblock {\url{https://doi.org/10.1093/mnras/stu199}}.

\bibitem[Abraham et~al.(2007)]{PierreAuger:2007pcg}
Abraham, J.;  Abreu, P.; Aglietta, M.; Aguirre, C.; Allard, D.; Allekotte, I.; Allen, J.; Allison, P.; Alvarez, C.; Alvarez-Muñiz, J.; et~al. [Pierre Auger Collaboration].
\newblock {Correlation of the highest energy cosmic rays with nearby
  extragalactic objects}.
\newblock {\em Science} {\bf 2007}, {\em 318},~938--943.
\newblock {\url{https://doi.org/10.1126/science.1151124}}.

\bibitem[Coleman et~al.(2023)Coleman, Eser, Mayotte, Sarazin, Schröder,
  Soldin, Venters, Aloisio, Alvarez-Muñiz, {Alves Batista}, Bergman, Bertaina,
  Caccianiga, Deligny, Dembinski, Denton, {di Matteo}, Globus, Glombitza,
  Golup, Haungs, Hörandel, Jaffe, Kelley, Krizmanic, Lu, Matthews, Mariş,
  Mussa, Oikonomou, Pierog, Santos, Tinyakov, Tsunesada, Unger, Yushkov,
  Albrow, Anchordoqui, Andeen, Arnone, Barghini, Bechtol, Bellido, Casolino,
  Castellina, Cazon, Conceição, Cremonini, Dujmovic, Engel, Farrar, Fenu,
  Ferrarese, Fujii, Gardiol, Gritsevich, Homola, Huege, Kampert, Kang, Kido,
  Klimov, Kotera, Kozelov, Leszczyńska, Madsen, Marcelli, Marisaldi,
  Martineau-Huynh, Mayotte, Mulrey, Murase, Muzio, Ogio, Olinto, Onel, Paul,
  Piotrowski, Plum, Pont, Reininghaus, Riedel, Riehn, Roth, Sako, Schlüter,
  Shoemaker, Sidhu, Sidelnik, Timmermans, Tkachenko, Veberic, Verpoest, Verzi,
  Vícha, Winn, Zas, and Zotov]{COLEMAN2023102819}
Coleman, A.; Eser, J.; Mayotte, E.; Sarazin, F.; Schröder, F.; Soldin, D.;
  Venters, T.; Aloisio, R.; Alvarez-Muñiz, J.; {Alves Batista}, R.;  et~al.
\newblock Ultra high energy cosmic rays The intersection of the Cosmic and
  Energy Frontiers.
\newblock {\em Astropart. Phys.} {\bf 2023}, {\em 149},~102819.
\newblock {\url{https://doi.org/10.1016/j.astropartphys.2023.102819}}.

\bibitem[Aloisio(2023)]{aloisio2023ultra}
Aloisio, R.
\newblock Ultra High Energy Cosmic Rays an overview.
\newblock \emph{Proc. J. Physics Conf. Ser.}  \textbf{2023},  \emph{2429}, 012008.

\bibitem[Caccianiga et~al.(2023)Caccianiga, Anchordoqui, Bianciotto, Bister,
  Bwembya, de~Jong, Eman, Falcke, Fodran, Galea, et~al.]{caccianiga2023update}
Caccianiga, L.; Anchordoqui, L.; Bianciotto, M.; Bister, T.; Bwembya, A.;
  de~Jong, S.; Eman, M.; Falcke, H.; Fodran, T.; Galea, C.;  et~al.
\newblock {Update on the searches for anisotropies in UHECR arrival directions
  with the Pierre Auger Observatory and the Telescope Array}.
\newblock {\em PoS} {\bf 2023}, {\em ICRC2023},~521.
\newblock {\url{https://doi.org/10.22323/1.444.0521}}.

\bibitem[Abdul~Halim et~al.(2023)Abdul~Halim, Abreu, Aglietta, Allekotte,
  Almeida~Cheminant, Almela, Aloisio, Alvarez-Muniz, Ammerman~Yebra, Anastasi,
  Anchordoqui, Andrada, Andringa, Aramo, Araújo~Ferreira, Arnone,
  Arteaga~Velazquez, Asorey, Assis, Avila, Avocone, Badescu, Bakalova,
  Balaceanu, Barbato, Bartz~Mocellin, Bellido, Berat, Bertaina, Bhatta,
  Bianciotto, Biermann, Binet, Bismark, Bister, Biteau, Blazek, Bleve, Blümer,
  Bohacova, Boncioli, Bonifazi, Bonneau~Arbeletche, Borodai, Brack,
  Brichetto~Orchera, Briechle, Bueno, Buitink, Buscemi, Büsken, Bwembya,
  Caballero-Mora, Cabana-Freire, Caccianiga, Caracas, Caruso, Castellina,
  Catalani, Cataldi, Cazon, Cerda, Cermenati, Chinellato, Chudoba, Chytka,
  Clay, Cobos~Cerutti, Colalillo, Coleman, Coluccia, Conceição, Condorelli,
  Consolati, Conte, Convenga, Correia~dos Santos, Costa, Covault, Cristinziani,
  Cruz~Sanchez, Dasso, Daumiller, Dawson, de~Almeida, de~Jesus, de~Jong,
  de~Mello~Neto, De~Mitri, de~Oliveira, de~Oliveira~Franco, de~Palma, de~Souza,
  De~Vito, Del~Popolo, Deligny, Denner, Deval, di~Matteo, Dobre, Dobrigkeit,
  D'Olivo, Domingues~Mendes, dos Anjos, dos Anjos, Ebr, Ellwanger, Emam, Engel,
  Epicoco, Erdmann, Etchegoyen, Evoli, Falcke, Farmer, Farrar, Fauth, Fazzini,
  Feldbusch, Fenu, Fernandes, Fick, Figueira, Filipcic, Fitoussi, Flaggs,
  Fodran, Fujii, Fuster, Galea, Galelli, García, Gaudu, Gemmeke, Gesualdi,
  Gherghel-Lascu, Ghia, GIACCARI, Giammarchi, Glombitza, Gobbi, Gollan, Golup,
  Gómez~Berisso, Gómez~Vitale, Gongora, González, Gonzalez, Goos, Gora,
  Gorgi, Gottowik, Grubb, Guarino, Guedes, Guido, Hahn, Hamal, Hampel, Hansen,
  Harari, Harvey, Haungs, Hebbeker, Hojvat, Hörandel, Horvath, Hrabovsky,
  Huege, Insolia, Isar, Janecek, Johnsen, Jurysek, Kääpä, Kampert,
  Keilhauer, Khakurdikar, Kizakke~Covilakam, Klages, Kleifges, Knapp, Kunka,
  Lago, Langner, Leigui~de Oliveira, Lema-Capeans, Lenok, Letessier-Selvon,
  Lhenry-Yvon, Lo~Presti, LOPES, Lu, Luce, Lundquist, Machado~Payeras,
  Majercakova, Mandat, Manning, Mantsch, Marafico, Mariani, Mariazzi, Maris,
  Marsella, Martello, Martinelli, Martínez~Bravo, Martins, Mastrodicasa,
  Mathes, Matthews, Matthiae, Mayotte, Mayotte, Mazur, Medina-Tanco, Meinert,
  Melo, Menshikov, Merx, Michal, Micheletti, Miramonti, Mollerach, Montanet,
  Morejon, Morello, Müller, Mulrey, Mussa, Muzio, Namasaka, Negi, Nellen,
  Nguyen, Nicora, Niculescu-Oglinzanu, Niechciol, Nitz, Nosek, Novotný, Nozka,
  Nucita, Nunez, Oliveira, Palatka, Pallotta, Panja, Parente, Paulsen,
  Pawlowsky, Pech, Pękala, Pelayo, Pereira, Pereira~Martins, Perez~Armand,
  Pérez~Bertolli, Perrone, Petrera, Petrucci, Pierog, Pimenta, Platino, Pont,
  Pothast, Pourmohammad~Shahvar, Privitera, Prouza, Puyleart, Querchfeld,
  Rautenberg, Ravignani, Reininghaus, Ridky, Riehn, Risse, Rizi, Rodrigues~de
  Carvalho, Rodriguez, Rodriguez~Rojo, Roncoroni, Rossoni, Roth, Roulet,
  Rovero, Ruehl, Saftoiu, Saharan, Salamida, Salazar, Salina, Sanabria~Gomez,
  Sánchez, Santos, Santos, Sarazin, Sarmento, Sato, Savina, Schäfer,
  Scherini, Schieler, Schimassek, Schimp, Schlüter, Schmidt, Scholten,
  Schoorlemmer, Schovanek, Schröder, Schulte, Schulz, Sciutto, Scornavacche,
  Segreto, Sehgal, Shivashankara, Sigl, Silli, Sima, Simon, Smau, Smida,
  Sommers, Soriano, Squartini, Stadelmaier, Stanca, Stanič, Stasielak, Stassi,
  Strähnz, Straub, Suárez-Durán, Suomijarvi, Supanitsky, Svozilikova,
  Szadkowski, Tapia, Taricco, Timmermans, Tkachenko, Tobiska, Todero~Peixoto,
  Tomé, Torrès, Travaini, Travnicek, Trimarelli, Tueros, Unger, Vaclavek,
  Vacula, Valdés~Galicia, Valore, Varela, Vásquez-Ramírez, Veberic, Ventura,
  Vergara~Quispe, Verzi, Vicha, Vink, Vlastimil, Vorobiov, Watanabe, Watson,
  Weindl, Wiencke, Wilczyński, Wittkowski, Wundheiler, Yue, Yushkov,
  Zapparrata, Zas, Zavrtanik, and Zavrtanik]{AbdulHalim:20232A}
Abdul~Halim, A.; Abreu, P.; Aglietta, M.; Allekotte, I.; Almeida~Cheminant, K.;
  Almela, A.; Aloisio, R.; Alvarez-Muniz, J.; Ammerman~Yebra, J.; Anastasi,
  G.A.;  et~al.
\newblock {Highlights from the Pierre Auger Observatory}.
\newblock {\em PoS} {\bf 2023}, {\em ICRC2023},~016.
\newblock {\url{https://doi.org/10.22323/1.444.0016}}.

\bibitem[Anchordoqui(2019)]{ANCHORDOQUI20191}
Anchordoqui, L.A.
\newblock Ultra-high-energy cosmic rays.
\newblock {\em Phys. Rep.} {\bf 2019}, {\em 801},~1--93.
  {\url{https://doi.org/10.1016/j.physrep.2019.01.002}}.

\bibitem[Bergman(2007)]{BERGMAN200719}
Bergman, D.
\newblock Observation of the GZK Cutoff Using the HiRes Detector.
\newblock {\em Nucl. Phys.  Proc. Suppl.} {\bf 2007}, {\em
  165},~19--26.
  {\url{https://doi.org/10.1016/j.nuclphysbps.2006.11.004}}.

\bibitem[Aab et~al.(2017)Aab, Abreu, Aglietta, Al~Samarai, Albuquerque,
  Allekotte, Almela, Castillo, Alvarez-Mu{\~n}iz, Anastasi,
  et~al.]{aab2017combined}
Aab, A.; Abreu, P.; Aglietta, M.; Al~Samarai, I.; Albuquerque, I.; Allekotte,
  I.; Almela, A.; Castillo, J.A.; Alvarez-Mu{\~n}iz, J.; Anastasi, G.;  et~al.
\newblock Combined fit of spectrum and composition data as measured by the
  Pierre Auger Observatory.
\newblock {\em J. Cosmol. Astropart. Phys.} {\bf 2017}, {\em
  2017},~038.

\bibitem[Fargion et~al.(2023)Fargion, {De Sanctis Lucentini}, and
  Khlopov]{FargionICRC2023PoS395}
Fargion, D.; {De Sanctis Lucentini}, P.G.; Khlopov, M.Y.
\newblock {UHECR multiplets versus Hot spots clustering in Auger sky: The
  forgotten signals}.
\newblock {\em PoS} {\bf 2023}, {\em ICRC2023},~395.
\newblock {\url{https://doi.org/10.22323/1.444.395}}.

\bibitem[Taylor et~al.(2023)Taylor, Matthews, and Bell]{Taylor:2023qdy}
Taylor, A.M.; Matthews, J.H.; Bell, A.R.
\newblock {UHECR echoes from the Council of Giants}.
\newblock {\em Mon. Not. Roy. Astron. Soc.} {\bf 2023}, {\em 524},~631--642.
\newblock {\url{https://doi.org/10.1093/mnras/stad1716}}.

\bibitem[Biteau(2021)]{Biteau:2021pru}
Biteau, J.
\newblock {Stellar Mass and Star Formation Rate within a Billion Light-years}.
\newblock {\em Astrophys. J. Supp.} {\bf 2021}, {\em 256},~15.
\newblock {\url{https://doi.org/10.3847/1538-4365/ac09f5}}.

\bibitem[Fargion(2011)]{FARGION2011111}
Fargion, D.
\newblock Coherent and random UHECR spectroscopy of lightest nuclei along Cen
  A: Shadows on GZK tau neutrinos spread in a near sky and time.
\newblock {\em Nucl. Instruments Methods Phys. Res. Sect. Accel. Spectrometers Detect. Assoc. Equip.} {\bf 2011},
  {\em 630},~111--114.
  {\url{https://doi.org/10.1016/j.nima.2010.06.040}}.




\bibitem[Abreu et~al.(2012)Abreu, Aglietta, Ahn, Albuquerque, Allard,
  Allekotte, Allen, Allison, Castillo, Alvarez-Mu{\~n}iz,
  et~al.]{abreu2012search}
Abreu, P.; Aglietta, M.; Ahn, E.; Albuquerque, I.F.d.M.; Allard, D.; Allekotte,
  I.; Allen, J.; Allison, P.; Castillo, J.A.; Alvarez-Mu{\~n}iz, J.;  et~al.
\newblock Search for signatures of magnetically-induced alignment in the
  arrival directions measured by the Pierre Auger Observatory.
\newblock {\em Astropart. Phys.} {\bf 2012}, {\em 35},~354--361.
\newblock {\url{https://doi.org/10.1016/j.astropartphys.2011.10.004}}.

\bibitem[Fargion et~al.(2023)Fargion, De~Sanctis~Lucentini, and
  Khlopov]{Fargion:2023yiy}
Fargion, D.; De~Sanctis~Lucentini, P.G.; Khlopov, M.Y.
\newblock UHECR Signatures and Sources.
\newblock {\em EPJ Web Conf.} {\bf 2023}, {\em 283},~04010.
\newblock {\url{https://doi.org/10.1051/epjconf/202328304010}}.



\bibitem[Fargion et~al.(1999)Fargion, Mele, and Salis]{fargion1999ultra}
Fargion, D.; Mele, B.; Salis, A.
\newblock Ultra-high-energy neutrino scattering onto relic light neutrinos in
  the galactic halo as a possible source of the highest energy extragalactic
  cosmic rays.
\newblock {\em  Astrophys. J.} {\bf 1999}, {\em 517},~725.

\bibitem[Weiler(1999)]{Weiler:1997sh}
Weiler, T.J.
\newblock {Cosmic ray neutrino annihilation on relic neutrinos revisited: A
  Mechanism for generating air showers above the Greisen-Zatsepin-Kuzmin
  cutoff}.
\newblock {\em Astropart. Phys.} {\bf 1999}, {\em 11},~303--316.
\newblock {\url{https://doi.org/10.1016/S0927-6505(98)00068-1}}.

\bibitem[Fargion and Salis(1998)]{fargion1998inverse}
Fargion, D.; Salis, A.
\newblock Inverse Compton scattering off black body radiation in high energy
  physics and gamma (MeV--TeV) astrophysics.
\newblock {\em Physics-Uspekhi} {\bf 1998}, {\em 41},~823.

\bibitem[Piazzoli et~al.(2022)Piazzoli, Liu, della Volpe, Cao, Chiavassa,
  Piazzoli, Guo, Ksenofontov, Martineau-Huynh, Martraire, Ma, Ma, Stenkin,
  Yuan, Zeng, Zhang, Zhang, and Zhu]{Piazzoli_2022}
Piazzoli, B.D.; Liu, S.M.; della Volpe, D.; Cao, Z.; Chiavassa, A.; Piazzoli,
  B.D.; Guo, Y.Q.; Ksenofontov, L.T.; Martineau-Huynh, O.; Martraire, D.;
  et~al.
\newblock Chapter 4 Cosmic-Ray Physics *.
\newblock {\em Chin. Phys. C} {\bf 2022}, {\em 46},~030004.
\newblock {\url{https://doi.org/10.1088/1674-1137/ac3faa}}.

\bibitem[Watson(2023)]{Watson:2023uko}
Watson, A.A.
\newblock {A Brief History of the Study of High Energy Cosmic Rays using Arrays
  of Surface Detectors}.
\newblock {\em EPJ Web Conf.} {\bf 2023}, {\em 283},~01002.
\newblock {\url{https://doi.org/10.1051/epjconf/202328301002}}.

\bibitem[Huchra et~al.(2012)]{huchra20122mass}
Huchra, J.P.;  Macri, L.M.; Masters, K.L.; Jarrett, T.H.; Berlind, P.; Calkins, M.; Crook, A.C.; Cutri, R.; Erdoğdu, P.; Falco, E.; et~al.
\newblock The 2MASS Redshift Survey—Description and data release.
\newblock {\em  Astrophys. J. Suppl. Ser.} {\bf 2012}, {\em
  199},~26.
\newblock {\url{https://doi.org/10.1088/0067-0049/199/2/26}}.


\bibitem[Morejon et~al.(2019)Morejon, Fedynitch, Boncioli, Biehl, and
  Winter]{Morejon:2019pfu}
Morejon, L.; Fedynitch, A.; Boncioli, D.; Biehl, D.; Winter, W.
\newblock {Improved photomeson model for interactions of cosmic ray nuclei}.
\newblock {\em JCAP} {\bf 2019}, {\em 11},~007.
\newblock {\url{https://doi.org/10.1088/1475-7516/2019/11/007}}.

\bibitem[Fujii(2024)]{Fujii:2024sys}
Fujii, T.
\newblock {Rapporteur Talk: CRI}.
\newblock {\em PoS} {\bf 2024}, {\em ICRC2023},~031.
\newblock {\url{https://doi.org/10.22323/1.444.0031}}.

\bibitem[Abbasi et~al.(2023)]{TelescopeArray:2023sbd}
Abbasi, R.U.;Allen, M.G.; Arimura, R.; Belz, J.W.; Bergman, D.R.; Blake, S.A.; Shin, B.K.; Buckland, I.J.; Cheon, B.G.; Fujii, T.; et~al. [Telescope Array Collaboration]
\newblock {An extremely energetic cosmic ray observed by a surface detector
  array}.
\newblock {\em Science} {\bf 2023}, {\em 382},~903--907.
\newblock {\url{https://doi.org/10.1126/science.abo5095}}.




\bibitem[Kashlinsky et~al.(2024)Kashlinsky, Atrio-Barandela, and
  Shrader]{Kashlinsky_2024}
Kashlinsky, A.; Atrio-Barandela, F.; Shrader, C.S. 
\newblock Probing the Dipole of the Diffuse Gamma-Ray Background.
\newblock {\em  Astrophys. J. Lett.} {\bf 2024}, {\em 961},~L1.
\newblock {\url{https://doi.org/10.3847/2041-8213/acfedd}}.



\bibitem[{Bird} et~al.(1995){Bird}, {Corbato}, {Dai}, {Elbert}, {Green},
  {Huang}, {Kieda}, {Ko}, {Larsen}, {Loh}, {Luo}, {Salamon}, {Smith},
  {Sokolsky}, {Sommers}, {Tang}, and {Thomas}]{1995ApJ...441..144B}
{Bird}, D.J.; {Corbato}, S.C.; {Dai}, H.Y.; {Elbert}, J.W.; {Green}, K.D.;
  {Huang}, M.A.; {Kieda}, D.B.; {Ko}, S.; {Larsen}, C.G.; {Loh}, E.C.;  et~al.
\newblock {Detection of a Cosmic Ray with Measured Energy Well beyond the
  Expected Spectral Cutoff due to Cosmic Microwave Radiation}.
\newblock {\em Apj} {\bf 1995}, {\em 441},~144.
\newblock {\url{https://doi.org/10.1086/175344}}.


\bibitem[Fargion(2002)]{fargion2002discovering}
Fargion, D.
\newblock Discovering ultra-high-energy neutrinos through horizontal and upward
  $\tau$ air showers: evidence in terrestrial gamma flashes?
\newblock {\em  Astrophys. J.} {\bf 2002}, {\em 570},~909.

\bibitem[Mayotte and Fitoussi(2018)]{Mayotte2022Aq}
Mayotte, E.; Fitoussi, T.
\newblock Update on the indication of a mass-dependent anisotropy above
  1e18.7eV in the hybrid data of the Pierre Auger Observatory. \emph{European PhysicsJ Web Conf. \textbf{2022}, \emph{2023}, 03003.}

\bibitem[Aab et~al.(2020)Aab, Abreu, Aglietta, Albury, Allekotte, Almela,
  Castillo, Alvarez-Muñiz, Batista, Anastasi, Anchordoqui, Andrada, Andringa,
  Aramo, Ferreira, Asorey, Assis, Avila, Badescu, Bakalova, Balaceanu, Barbato,
  Luz, Becker, Bellido, Berat, Bertaina, Bertou, Biermann, Bister, Biteau,
  Blanco, Blazek, Bleve, Boháčová, Boncioli, Bonifazi, Arbeletche, Borodai,
  Botti, Brack, Bretz, Briechle, Buchholz, Bueno, Buitink, Buscemi,
  Caballero-Mora, Caccianiga, Calcagni, Cancio, Canfora, Caracas, Carceller,
  Caruso, Castellina, Catalani, Cataldi, Cazon, Cerda, Chinellato, Choi,
  Chudoba, Chytka, Clay, Cerutti, Colalillo, Coleman, Coluccia, Conceição,
  Condorelli, Consolati, Contreras, Convenga, Covault, Dasso, Daumiller,
  Dawson, Day, de~Almeida, de~Jesús, de~Jong, Mauro, de~Mello~Neto, Mitri,
  de~Oliveira, de~Oliveira~Franco, de~Souza, Debatin, del Río, Deligny,
  Dhital, Matteo, Castro, Dobrigkeit, D'Olivo, Dorosti, dos Anjos, Dova, Ebr,
  Engel, Epicoco, Erdmann, Escobar, Etchegoyen, Falcke, Farmer, Farrar, Fauth,
  Fazzini, Feldbusch, Fenu, Fick, Figueira, Filipčič, Fodran, Freire, Fujii,
  Fuster, Galea, Galelli, García, Vegas, Gemmeke, Gesualdi, Gherghel-Lascu,
  Ghia, Giaccari, Giammarchi, Giller, Glombitza, Gobbi, Golup, Berisso, Vitale,
  Gongora, Goy, Hampel, Hansen, Harari, Harvey, Haungs, Hebbeker, Heck, Hill,
  Hojvat, Hörandel, Horvath, Hrabovský, Huege, Hulsman, Insolia, Isar,
  Johnsen, Jurysek, A.nzález, Goos, Góra, Gorgi, Gottowik, Grubb, Guarino,
  Guedes, Guido, Hahn, Kääpä, Kampert, Keilhauer, Kemp, Klages, Kleifges,
  Kleinfeller, Köpke, Mezek, Lago, LaHurd, Lang, de~Oliveira, Lenok,
  Letessier-Selvon, Lhenry-Yvon, Presti, Lopes, López, Lorek, Luce, Lucero,
  Payeras, Malacari, Mancarella, Mandat, Manning, Manshanden, Mantsch,
  Marafico, Mariazzi, Maris, Marsella, Martello, Martinez, Bravo, Mastrodicasa,
  Mathes, Matthews, Matthiae, Mayotte, Mazur, Medina-Tanco, Melo, Menshikov,
  Merenda, Michal, Micheletti, Miramonti, Mockler, Mollerach, Montanet,
  Morello, Mostafá, Müller, Muller, Mulrey, Mussa, Muzio, Namasaka, Nellen,
  Niculescu-Oglinzanu, Niechciol, Nitz, Nosek, Novotny, Nožka, Nucita,
  Núñez, Palatka, Pallotta, Panetta, Papenbreer, Parente, Parra, Pech,
  Pedreira, Pekala, Pelayo, Peña-Rodriguez, Armand, Perlin, Perrone, Peters,
  Petrera, Pierog, Pimenta, Pirronello, Platino, Pont, Pothast, Privitera,
  Prouza, Puyleart, Querchfeld, Rautenberg, Ravignani, Reininghaus, Ridky,
  Riehn, Risse, Ristori, Rizi, de~Carvalho, Rojo, Roncoroni, Roth, Roulet,
  Rovero, Ruehl, Saffi, Saftoiu, Salamida, Salazar, Salina, Gomez, Sánchez,
  Santos, Santos, Sarazin, Sarmento, Sarmiento-Cano, Sato, Savina, Schäfer,
  Scherini, Schieler, Schimassek, Schimp, Schlüter, Schmidt, Scholten,
  Schovánek, Schröder, Schröder, Sciutto, Scornavacche, Shellard, Sigl,
  Silli, Sima, Smída, Sommers, Soriano, Souchard, Squartini, Stadelmaier,
  Stanca, Stanič, Stasielak, Stassi, Streich, Suárez-Durán, Sudholz,
  Suomijärvi, Supanitsky, Supík, Szadkowski, Taboada, Tapia, Timmermans,
  Tobiska, Peixoto, Tomé, Elipe, Travaini, Travnicek, Trimarelli, Trini,
  Tueros, Ulrich, Unger, Urban, Vaclavek, Vacula, Galicia, Valiño, Valore, van
  Vliet, Varela, Cárdenas, Vásquez-Ramírez, Veberič, Ventura, Quispe,
  Verzi, Vicha, Villaseñor, Vink, Vorobiov, Wahlberg, Watson, Weber, Weindl,
  Wiencke, Wilczyński, Winchen, Wirtz, Wittkowski, Wundheiler, Yushkov,
  Zapparrata, Zas, Zavrtanik, Zavrtanik, Zehrer, Zepeda, Ziolkowski, and
  Zuccarello]{Aab_2020}
Aab, A.; Abreu, P.; Aglietta, M.; Albury, J.; Allekotte, I.; Almela, A.;
  Castillo, J.A.; Alvarez-Muñiz, J.; Batista, R.A.; Anastasi, G.;  et~al.
\newblock Search for magnetically-induced signatures in the arrival directions
  of ultra-high-energy cosmic rays measured at the Pierre Auger Observatory.
\newblock {\em J. Cosmol. Astropart. Phys.} {\bf 2020}, {\em
  2020},~017.
\newblock {\url{https://doi.org/10.1088/1475-7516/2020/06/017}}.

\bibitem[Fargion et~al.(2001)Fargion, Grossi, and {De Sanctis
  Lucentini}]{fargion2001clustering}
Fargion, D.; Grossi, M.; {De Sanctis Lucentini}, P.G.
\newblock Clustering, Anisotropy, Spectra of Ultra High Energy Cosmic Ray:
  Finger-Prints of Relic Neutrinos Masses in Dark Halos.
\newblock {\em J. Phys. Soc. Jpn.} {\bf 2001}, {\em
  70},~46--57.

\bibitem[Protheroe and Meyer(2000)]{protheroe2000infrared}
Protheroe, R.; Meyer, H.
\newblock An infrared background-TeV gamma-ray crisis?
\newblock {\em Phys. Lett. B} {\bf 2000}, {\em 493},~1--6.

\bibitem[Cao et~al.(2023)]{LHAASO:2023lkv}
Cao, Z.;  Aharonian, F.; An, Q.; Axikegu, A.; Bai, Y.X.; Bao, Y.W.; Bastieri, D.; Bi, X.J.; Bi, Y.J.; Cai, J.T.;  et~al. [LHAASO Collaboration]
\newblock {Very high energy gamma-ray emission beyond 10 TeV from GRB 221009A}.
\newblock {\em Sci. Adv.} {\bf 2023}, {\em 9},~adj2778.
\newblock {\url{https://doi.org/10.1126/sciadv.adj2778}}.

\bibitem[Fraija et~al.(2023)Fraija, Aguilar-Ruiz, Galv\'an, Onsurbe, and
  Dainotti]{Fraija:2023yil}
Fraija, N.; Aguilar-Ruiz, E.; Galv\'an, A.; Onsurbe, J.A.d.D.; Dainotti, M.G.
\newblock {The unprecedented flaring activities around Mrk 421 in 2012 and
  2013: The test for neutrino and UHECR event connection}.
\newblock {\em JHEAp} {\bf 2023}, {\em 40},~55--67.
\newblock {\url{https://doi.org/10.1016/j.jheap.2023.10.003}}.

\bibitem[Marafico et~al.(2024)Marafico, Biteau, Condorelli, Deligny, and
  Bregeon]{Marafico:2024qgh}
Marafico, S.; Biteau, J.; Condorelli, A.; Deligny, O.; Bregeon, J.
\newblock {Closing the net on transient sources of ultra-high-energy cosmic
  rays} \emph{arxiv}~{\bf 2024}, arxiv:2405.17179.

\bibitem[Abbasi et~al.(2024)]{TelescopeArray:2024oux}
Abbasi, R.U.; Abe, Y.; Abu-Zayyad, T.; Allen, M.; Arai, Y.; Arimura, R.; Barcikowski, E.; Belz, J.W.; Bergman, D.R.; Blake, S.A.;  et~al. [Telescope Array Collaboration]
\newblock {Isotropy of Cosmic Rays beyond 1020\,\,eV Favors Their Heavy Mass
  Composition}.
\newblock {\em Phys. Rev. Lett.} {\bf 2024}, {\em 133},~041001.
\newblock {\url{https://doi.org/10.1103/PhysRevLett.133.041001}}.

\bibitem[Abdul~Halim et~al.(2024)]{PierreAuger:2024hrj}
Abdul~Halim, A.;   Abreu, P.; Aglietta, M.; Allekotte, I.; Cheminant, K.A.; Almela, A.; Aloisio, R.; Alvarez-Muñiz, J.; Yebra, J.A.; Anastasi, G.A.;  et~al. [Pierre Auger Collaboration]
\newblock {The flux of ultra-high-energy cosmic rays along the supergalactic
  plane measured at the Pierre Auger Observatory} \emph{arxiv} {\bf 2024}, arxiv:2407.06874.

\end{thebibliography}
\end{document}